\documentclass[%
 reprint,
amsmath,amssymb,
 aps,
superscriptaddress]{revtex4-1}

\newcommand\Tstrut{\rule[4ex]{0pt}{0pt}}        
\newcommand\Bstrut{\rule[-3ex]{0pt}{0pt}}   
\usepackage{book tabs}
\usepackage{graphicx}
\usepackage{dcolumn}
\usepackage{bm}
\usepackage{relsize}
\usepackage{float}
\usepackage{adjustbox}
\usepackage{color}
\usepackage[normalem]{ulem}

\begin{document}

\preprint{APS/123-QED}

\title{Reconstructing Horndeski theories from phenomenological modified gravity and dark energy models on cosmological scales}

\author{Joe~Kennedy}
\affiliation{Institute for Astronomy, University of Edinburgh, Royal Observatory, \\ Blackford Hill, Edinburgh, EH9 3HJ, U.K.}
\author{Lucas~Lombriser}
\affiliation{Institute for Astronomy, University of Edinburgh, Royal Observatory, \\ Blackford Hill, Edinburgh, EH9 3HJ, U.K.}
\affiliation{D\'{e}partement de Physique Th\'{e}orique,
Universit\'{e} de Gen\`{e}ve,
24 quai Ernest Ansermet,
1211 Gen\`{e}ve 4,
Switzerland}
\author{Andy~Taylor}
\affiliation{Institute for Astronomy, University of Edinburgh, Royal Observatory, \\ Blackford Hill, Edinburgh, EH9 3HJ, U.K.}

\date{\today}

\begin{abstract}
Recently we have
derived a set of mapping relations that enables
the reconstruction of the family of Horndeski scalar-tensor theories which
reproduce the background dynamics and linear perturbations of a given set of effective field theory of dark energy coefficients.
In this paper we present a number of applications of this reconstruction.
We examine the form of the underlying theories behind different phenomenological parameterizations of modified gravity and dark energy used in the literature, as well as examine theories that exhibit weak gravity, linear shielding, and minimal self-acceleration.
Finally, we propose a new inherently stable parametrization basis for modified gravity and dark energy models.

\end{abstract}

\pacs{Valid PACS appear here}
\maketitle

\section{Introduction}
\label{sec:intro}
The observation of the late-time accelerated expansion of
our Universe~\cite{Riess:1998cb, Perlmutter:1998np} remains one of the greatest
puzzles
in physics. Owing to the large number of theories that have
been proposed as explanations for the accelerated expansion~\cite{Clifton:2011jh, Koyama:2015vza, Joyce:2016vqv, Joyce:2014kja}, efficient methods must be devised to narrow down the theory space. In doing so, one hopes to achieve a deeper understanding of the physical mechanism
driving the cosmic late-time expansion.
One of the simplest approaches to tackle the accelerated expansion, beyond a cosmological constant, is to assume that it is driven by the dynamics of a scalar field that acts on large scales. This scalar field could be the low-energy effective remnant from some more fundamental theory of gravity, the fine details of which are not relevant on the scales of interest.
When one adds a scalar field to
gravity it is necessary to do
so in such a way that it evades Ostrogradski instabilities. The most general
scalar-tensor action yielding up to second-order equations of motion was originally derived by Horndeski and independently rediscovered much later in a different context \cite{Horndeski:1974wa,Deffayet:2011gz,Kobayashi2011}.
More general, higher-order actions have then been devised that avoid Ostrogradski ghosts by the avoidance of the non-degeneracy condition~\cite{Zumalacarregui:2013pma, Gleyzes:2014dya, Langlois:2015cwa}.

The recent LIGO/Virgo measurement of the gravitational wave GW170817~\cite{TheLIGOScientific:2017qsa} emitted by a binary neutron star merger
with the simultaneous observations of electromagnetic counterparts~\cite{Monitor:2017mdv,GBM:2017lvd} has led to a significant reduction of the available theory space at late
times,
as was first anticipated in Refs.~\cite{Lombriser:2015sxa,Lombriser:2016yzn}.
The GW170817 event occurred in the NGC 4993 galaxy of the Hydra cluster at a distance of about 40~Mpc and enabled a constraint on the relative deviation of the speed of gravity $c_T$ from the speed of light ($c=1$) at $\mathcal{O}(10^{-15})$ for $z\lesssim0.01$~\cite{Monitor:2017mdv}.
This agrees with forecasts~\cite{Nishizawa:2014zna,Lombriser:2015sxa} inferred from the increased likelihood with increasing volume at the largest distances resolved by the detectors, expecting a few candidate events per year, and emission time uncertainties.
It was anticipated that
the measurement
would imply that a genuine cosmic self-acceleration from Horndeski scalar-tensor theory and its degenerate higher-order extensions, including the Galileon theories, can no longer arise from an evolving speed of gravity and must instead be attributed to 
a running effective Planck mass~\cite{Lombriser:2015sxa}.
The minimal evolution of the Planck mass required for self-acceleration with $c_T=1$ was derived in Ref.~\cite{Lombriser:2016yzn}
and
was shown to provide a $3\sigma$ worse fit to cosmological data than a cosmological constant.
Strictly speaking,
this only applies to Horndeski theories, where
$c_T=1$ breaks a fundamental degeneracy in the large-scale structure
produced by the theory space~\cite{Lombriser:2014ira,Lombriser:2015sxa}.
Generalizations of the Horndeski action reintroduce this degeneracy~\cite{Lombriser:2014ira}
but self-acceleration in general scalar-tensor theories is expected to be conclusively testable at the $5\sigma$ level with Standard Sirens~\cite{Lombriser:2015sxa} (also see Refs.~\cite{Saltas:2014dha,Nishizawa:2017nef,Belgacem:2017ihm,Amendola:2017ovw}), eventually allowing an extension of this No-Go result.
The minimal
model serves as a null-test for self-acceleration from modified gravity. It is therefore worth examining whether future observational probes of the large-scale structure are capable of tightening the constraint beyond the $3\sigma$-level.
Finally, the measurement of $c_T\simeq1$ with GW170817 in particular implies that the quintic and kinetically coupled quartic Horndeski Lagrangians must be negligible at late times~\cite{Kimura:2011qn} (also see e.g., Refs.~\cite{Jimenez:2015bwa,Lombriser:2015sxa,Brax:2015dma,Bettoni:2016mij,Pogosian:2016pwr,Creminelli:2017sry, Sakstein:2017xjx, Ezquiaga:2017ekz, Baker:2017hug, Crisostomi:2017pjs, Linder:2018jil} for more recent discussions).
The measurement also led to a range of further astrophysical and cosmological implications (see, e.g., Ref.~\cite{Wang:2017rpx}). \vspace{0.15cm}
\\
\indent Despite giving strong restrictions on the set of scalar-tensor theories that could explain the accelerated expansion, there remains a great deal of freedom in the model space after the GW170817 observation
and the phenomenological study of the quintic and kinetically coupled quartic Horndeski Lagrangians should not be dismissed so soon.
There are two important aspects to be considered in this argument.
On the one hand, the speeds of gravity and light are only constrained to be effectively equal at the low redshifts of $z\lesssim0.01$. This certainly applies to the regime of cosmic acceleration but not to the early Universe, where a decaying deviation in $c_T$ could still lead to observable signatures without invoking fine-tuning.
Moreover, for more general scalar-tensor theories, the linear shielding mechanism~\cite{Lombriser:2014ira} may be extended to a modified gravitational wave propagation, where the Horndeski terms could cause cosmic self-acceleration while other terms may come to dominate for the wavelengths relevant to GW170817.
These wavelengths differ by those associated with cosmic acceleration by $\mathcal{O}(10^{19})$ \cite{Battye:2018ssx}.
Hence, in this paper we will not exclusively restrict to the models satisfying the GW170817 constraint, envisaging more general applications of the methods presented.

To cover the vast landscape of dark energy and modified gravity models and compare predictions to observations, it is desirable to develop
efficient and systemic frameworks.
The effective field theory of dark energy (EFT) is such a tool~\cite{Creminelli:2008wc, Park:2010cw,
Gubitosi2012, Bloomfield:2012ff, Gleyzes:2013ooa, Tsujikawa:2014mba, Bellini:2014fua, Lagos:2016wyv}.
At the level of the background and linear perturbations Horndeski theory can be described by five free functions of time.
Despite the utility of EFT, by its construction it cannot give a full description of the underlying physical theory. 
Currently, one either has to start from a given fully covariant theory and compute the EFT coefficients in terms of the functions defined in the covariant action, or take a phenomenologically motivated parameterization for the EFT functions. In the first instance, one is essentially left with the original problem of having a large range of theories to compare with observations.
Following the second approach gives a general indication of the effects of modified gravity on different observational probes,
but it is generally unclear
what physical theories are being tested when a particular parameterization is adopted.

In a recent work~\cite{Kennedy:2017sof} we have developed
a mapping
from the EFT coefficients to the
family
of Horndeski models which
give rise to
the same background evolution and linear perturbations.
This mapping provides a method to study
the form of the Horndeski functions
determined from observations
on large scales.
One can furthermore address the question of what theories various phenomenological parameterizations of the EFT functions correspond to. 
This paper provides a number of applications of this reconstruction. For example, we examine the form of the underlying theories corresponding to two commonly used EFT parameterizations
for late-time modifications motivated by cosmic acceleration.
Reconstructed actions that exhibit minimal self-acceleration and linear shielding are also presented.
We furthermore apply the reconstruction to
phenomenological parameterizations such as a modified Poisson equation and gravitational slip~\cite{Uzan:2006mf, Amendola:2007rr, Caldwell:2007cw, Hu:2007pj, Zhang:2007nk} as well as the growth-index parametrization~\cite{PeeblesLSS, Wang:1998gt, Linder:2005in}.
These
are the primary parameters that the next generation of galaxy-redshift surveys will target~\cite{Laureijs:2011gra,Amendola:2012ys,Ivezic:2008fe}.
With the reconstruction it is possible to connect these parameterizations with viable covariant theories, and explore the region of the theory space being sampled when a particular parameterization is adopted.
The reconstruction is
also
applied
to a phenomenological model
that exhibits a weakening of the growth of structure relative to $\Lambda$CDM today,
which may be of interest to address potential observational tensions~\cite{Abbott:2017wau, Amon:2017lia}.
Finally, in every analysis of the EFT model space it is necessary to ensure that the chosen model parameters do not lead to ghost or gradient instabilities.
When comparing models with observations this can, for instance, lead to a highly inefficient sampling of the model space
and misleading statistical constraints due to complicated stability priors.
To avoid those issues, we propose an alternative parameterization of the EFT function space, which uses the stability parameters directly as the basis set
such that every sample drawn from that space is inherently stable.

The paper is organized as follows. In Sec.~\ref{StabCond} we 
briefly review the EFT formalism and specify
the stability criteria imposed on the model space.
We then propose our new inherently stable EFT basis that we argue is most suitable for statistical comparisons of the available theory space to observations.
Sec.~\ref{sec:results}
covers a number of different
reconstructions,
ranging from commonly adopted parameterizations encountered in the literature (Sec.~\ref{sec:IIIA}) to models
for
minimal self-acceleration (Sec.~\ref{sec:IIIB}), linear shielding (Sec.~\ref{sec:IIIC}),
phenomenological modifications of the Poisson equation and gravitational slip (Sec.~\ref{sec:IIID}), the growth-index parametrization (Sec.~\ref{sec:IIIE}), and
weak gravity (Sec.~\ref{sec:IIIF}).
In Sec.~\ref{sec:IIIG} we provide an example of a reconstruction from the inherently stable parameter space.
Finally, we discuss
conclusions in Sec.~\ref{sec:conclusions} and
inspect
the impact of the choice of EFT parametrization on the reconstructed theories in the Appendix.

%%%% Different Parameterizations %%%%

\section{Horndeski gravity and Effective field theory formalism}
\label{StabCond}
The most general local four-dimensional scalar-tensor theory evading Ostrogradski instabilities and restricted to at most second-order equations of motion 
is given by the Horndeski action~\cite{Horndeski:1974wa, Deffayet:2011gz, Kobayashi2011}
\begin{equation}
S=\sum_{i=2}^{5} \int d^{4} x \sqrt{-g} \, \mathcal{L}_{i} \,,
\label{eq:Horndeski}
\end{equation}
where each $\mathcal{L}_{i}$ is given by
\begin{eqnarray}
\mathcal{L}_{2} & \equiv & G_{2}(\phi,X) \,, \label{eq:HorndeskiL2} \\
\mathcal{L}_{3} & \equiv & G_{3}(\phi, X)\Box \phi \,, \label{eq:HorndeskiL3} \\
\mathcal{L}_{4} & \equiv & G_{4}(\phi, X)R \nonumber\\
 & & -2G_{4X}(\phi, X) \left[(\Box \phi)^{2}-(\nabla^{\mu}\nabla^{\nu}\phi)(\nabla_{\mu}\nabla_{\nu}\phi)    \right] \,, \label{eq:g4} \\
\mathcal{L}_{5} & \equiv & G_{5}(\phi, X)G_{\mu \nu}\nabla^{\mu}\nabla^{\nu}\phi \nonumber\\
 & & +\frac{1}{3}G_{5X}(\phi, X) \left[(\Box \phi)^{3} -3(\Box \phi) (\nabla_{\mu}\nabla_{\nu}\phi)(\nabla^{\mu}\nabla^{\nu}\phi) \right. \nonumber\\
 & & \left. +2(\nabla_{\mu}\nabla_{\nu}\phi)(\nabla^{\sigma}\nabla^{\nu}\phi)(\nabla_{\sigma}\nabla^{\mu}\phi)   \right] \, , \label{eq:g5}
\end{eqnarray}
and $X \equiv g^{\mu\nu}\partial_{\mu}\phi\partial_{\nu}\phi$. 
Note that the GW170817 result $c_T\simeq1$ implies that $G_{4X} \simeq G_5 \simeq 0$~\cite{Kimura:2011qn} at redshifts $z\lesssim0.01$.
Throughout we adopt the flat Friedmann-Lema\^itre-Robertson-Walker (FLRW) metric for the background 
\begin{equation}
ds^{2}=-dt^{2}+a^{2}(t)d \textbf{x}^{2} \,,
\label{eq:lineelement}
\end{equation}
which describes a statistically spatially homogeneous and isotropic Universe. The scale factor $a(t)$ is normalized such that it equals one today.

The EFT formalism involves breaking time diffeomorphism invariance by adopting the unitary gauge where the scalar field is set equal to a time-like, monotonic function of time $\phi(t)$.
Specifically we choose the value of the scalar field to correspond to constant time hypersurfaces such that
\begin{equation}
\phi=t M_{*}^{2} \,,
\end{equation}
where $M_*$ is the bare Planck mass.
The broken time diffeomorphism invariance implies that the only terms which are allowed in the EFT action are those which are invariant under spatial diffeomorphisms with free time-dependent coefficients.
The allowed operators which are sufficient to describe Horndeski theory 
are the time-time component of the metric $g^{00}$ as well as the extrinsic curvature of the space-like hypersurfaces $K_{\mu\nu}=h_{\mu}^{\sigma}\nabla_{\sigma}n_{\nu}$, where the induced metric is $h_{\mu\nu}=g_{\mu\nu}+n_{\mu}n_{\nu}$ and $n_{\mu}$ is a time-like normal vector to the hypersurface. The last allowed operator is the three dimensional Ricci tensor $R^{(3)}_{\mu\nu}$, defined in the same way as the full Ricci tensor $R_{\mu\nu}$ but using $h_{\mu\nu}$ in place of $g_{\mu\nu}$.

At second order, Horndeski gravity corresponds to the EFT action \cite{Gubitosi2012, Bloomfield:2012ff, Gleyzes:2013ooa, Bloomfield:2013efa, Kennedy:2017sof}
\begin{align}
S = & \: S^{(0,1)}+S^{(2)}+S_{M}[g_{\mu\nu},\Psi_m] \,,
\label{eftlag}\\
S^{(0,1)} = & \: \frac{M_{*}^{2}}{2}\int d^{4}x \sqrt{-g} \left[ \Omega(t) R -2\Lambda(t)-\Gamma(t)\delta g^{00} \right] \,,
\label{eq:s01} \\
S^{(2)} = & \int d^{4}x \sqrt{-g} \left[ \frac{1}{2}M^{4}_2(t)(\delta g^{00})^2-\frac{1}{2}\bar{M}^{3}_{1}(t) \delta K \delta g^{00}\right. \nonumber\\
 & \left.-\bar{M}^{2}_{2}(t) \left( \delta K^2-\delta  K^{\mu\nu} \delta K_{\mu\nu} - \frac{1}{2} \delta R^{(3)}\delta g^{00} \right) \right] \,,
\label{eq:s2}
\end{align}
where $S^{(0,1)}$ describes the cosmological background evolution and $S^{(2)}$ describes the linear perturbations around it.
In general, various subsets of Horndeski theory lead to separate contributions from the EFT coefficients.
In particular, theories compatible with the GW1710817 observation must satisfy $\bar{M}^2_2(t)\simeq0$ at $z\lesssim0.01$.
Taking into account the Hubble expansion $H(t)\equiv\dot{a}/a$ and the two constraints from the Friedmann equations, Eqs.~\eqref{eftlag}--\eqref{eq:s2} contain five independent functions capable of describing the background and linear perturbations of Horndeski theory. 

One can also define an alternative basis for the EFT functions with a more direct physical interpretation~\cite{Bellini:2014fua}. See Table~I of Ref.~\cite{Gleyzes:2014rba} and Table~II of Ref.~\cite{Kennedy:2017sof} for the connection between the two descriptions, although bear in mind the different conventions. This basis is defined via 
\begin{eqnarray}
\alpha_{M} & \equiv & \frac{M_{*}^{2}\Omega^{\prime}+2(\bar{M}_{2}^{2})^{\prime}}{M_{*}^{2}\Omega+2\bar{M}_{2}^{2}} \, , \label{Am} \\
\alpha_{B} & \equiv & \frac{M_{*}^{2}H\Omega^{\prime}+\bar{M}_{1}^{3}}{2H\left( M_{*}^{2}\Omega+2\bar{M}_{2}^{2}     \right)} \, , \\
\alpha_{K} & \equiv & \frac{M_{*}^{2}\Gamma+4M_{2}^{4}}{H^{2}\left(M_{*}^{2}\Omega+2\bar{M}_{2}^{2} \right)} \, , \\
\alpha_{T} & \equiv & -\frac{2\bar{M}_{2}^{2}}{M_{*}^{2}\Omega+2\bar{M}_{2}^{2}} \label{At} \,,
\end{eqnarray}
where throughout this section primes denote derivatives with respect to $\ln a$. Here $\alpha_{M}$ describes the running of the effective Planck mass $M=\sqrt{M_{*}^{2}\Omega+2\bar{M}_{2}^{2}}$ defined through $\alpha_{M}=d \,  \textnormal{ln} \, M^{2} / d \, \textnormal{ln} \, a$, allowing for some variation in the strength of the gravitational coupling over time.
The function $\alpha_{B}$ describes a braiding or mixing between the kinetic contributions of the scalar and tensor fields.
The function $\alpha_{K}$ enters through the kinetic term of the scalar field and only becomes relevant on scales comparable to the horizon.
Finally, $\alpha_{T}$ describes the deviation of the speed of gravitational waves from the speed of light with $c_T^2=1+\alpha_T$.

\subsection{Stability Criteria}
\label{sec:IIA}

To ensure the absence of ghost and gradient instabilities it is necessary to impose certain constraints on the EFT functions.
For instance, in order to avoid a kinetic term with the wrong sign or an imaginary sound speed for the scalar modes one must have~\cite{Bellini:2014fua}
\begin{equation}
\alpha \equiv \alpha_{K}+6\alpha_{B}^{2} > 0 \,,  \quad c_{s}^{2}>0 \,,
\label{Stabcriteria1}
\end{equation}
where the soundspeed $c_s$ is given by
\begin{align}
c_{s}^{2}=&-\frac{2}{\alpha}\left[\alpha_{B}^{\prime}+(1+\alpha_{T})(1+\alpha_{B})^{2} \right. \nonumber \\
 &\left. -\left(1+\alpha_{M}-\frac{H^{\prime}}{H} \right)(1+\alpha_{B})+\frac{\rho_{m}}{2H^{2}M^{2}}   \right] \,.
 \label{soundspeed}
\end{align}
Furthermore, the stability of the background to tensor modes requires
\begin{equation}
 c_T^2 > 0 \,, \quad M^2>0 \,.
 \label{Stabcriteria2}
\end{equation}
One must be careful when using parametrizations of the EFT functions to reconstruct covariant theories
that the stability conditions are satisfied. A way to achieve this that we adopt
in Secs.~\ref{sec:IIIA}--\ref{sec:IIIE}
is to set the soundspeed equal to unity and use this as a constraint on the EFT coefficients. 
It then remains to check that the other conditions are also satisfied by hand. This is somewhat restrictive as there are many viable stable scalar-tensor theories that do not have $c_{s}^{2}=1$.
An alternative approach is to directly parameterize the stability conditions as a new set of EFT functions (Secs.~\ref{sec:stableparam} and \ref{sec:IIIG}).

\subsection{A New Inherently Stable Parameterization} \label{sec:stableparam}

For generic tests of modified gravity and dark energy, a range of different time parametrizations (see Sec.~\ref{sec:IIIA}) are commonly adopted for the EFT coefficients in 
$S^{(0,1)}$ and $S^{(2)}$
or for the $\alpha_i$ functions.
These parameterizations do not a priori satisfy the stability criteria in Eqs.~\eqref{Stabcriteria1} and \eqref{Stabcriteria2}.
As a consequence the sampling in this parametrization, for example when conducting a Markov Chain Monte Carlo (MCMC) analysis to constrain the EFT parameter space with observations, can be highly inefficient. Only a very small fraction of the samples will hit a stable region of parameter space.
Moreover, the stability criteria can yield contours on the parameter space that are statistically difficult to interpret.
For instance, $\Lambda$CDM can be confined to a
narrow
corner of two intersecting edges produced by the stability requirements.
This corner may only be sparsely sampled and could lead to spurious evidence against concordance cosmology.

To avoid those issues, we propose here a new basis
for the parametrization of modified gravity and dark energy models in the effective field theory formalism.
We will make use of the GW170817 constraint $\alpha_T\simeq0$ at $z\lesssim0.01$ and assume that it applies throughout the late-time domain of interest here.
We define a function $B$ through
\begin{equation}
1+\alpha_B\equiv\frac{B'}{B} \,.
\end{equation}
Eq.~\eqref{soundspeed} can then be expressed as a linear homogeneous second-order differential equation
for $B$ with
\begin{equation}
 B'' - \left(1+\alpha_M-\frac{H'}{H} \right) B' + \left(\frac{\rho_m}{2H^2 M^2} + \frac{\alpha \: c_{s}^{2}}{2} \right) B = 0 \,.
\label{secondorderODE}
\end{equation}
By the existence and uniqueness theorem for ordinary differential equations a real solution exists for real boundary conditions on $B$ and $B'$.
Alternatively, we may provide an initial or present value $\alpha_{Bi}$ or $\alpha_{B0}$, respectively.

Hence an inherently stable parametrization of the EFT function space for modified gravity and dark energy can be defined by parametrizations of the basis %base
\begin{equation}
 M^2 > 0 \,, \quad c_s^2 > 0 \,, \quad \alpha > 0 \,, \quad \alpha_{B0} = \textrm{const.} \, ,
\end{equation}
along with the Hubble parameter $H$. The braiding function $\alpha_B$ can be determined from the integration of Eq.~\eqref{secondorderODE} and $\alpha_K$ from $\alpha_B$ and $\alpha$.

We advocate that this basis should be used for observational constraints on the EFT function space to avoid the problems described earlier.
It also provides a direct physical interpretation of the observational constraints.
While parametrizations in $H$ classify quintessence dark energy models where $\alpha>0$, $c_s^2$ describes more exotic dark energy models with $\alpha_{B0}\neq0$ adding an imperfection to the fluid and $M\neq M_*$ modifying gravity.
In $\Lambda$CDM, $M=M_*$, $c_s$ drops out and the remaining parameters vanish. 
This parameterization furthermore addresses the measure problem on the parameter space. While it is difficult to know a priori what is a reasonable prior range to place on the $\alpha_{i}$ parameters, it is much clearer in this physical parameterization. In addition, if measurements of these physical parameters seem to approach a fixed value it becomes easier to place bounds on the desired accuracy. 
We shall apply the reconstruction to a model defined in this basis in Sec.~\ref{sec:IIIG}.
Finally, note that one can easily add the beyond-Horndeski parameter $\alpha_H$ to this basis, which will introduce a modification in $c_s^2$.

%%%%%%%% RESULTS %%%%%%%%

\section{Reconstructing covariant theories}
\label{sec:results}

We now present a series of applications of the
mapping relations derived
in Ref.~\cite{Kennedy:2017sof}. We begin with a reconstruction of common parameterizations of the EFT functions used in the literature (Sec.~\ref{sec:IIIA}) and then examine the form of the underlying theory of the minimal self-acceleration model (Sec.~\ref{sec:IIIB}) and theories that exhibit linear shielding (Sec.~\ref{sec:IIIC}).
Following this, we discuss reconstructions from more phenomenological modifications of gravity with a modified Poisson equation and a gravitational slip (Sec.~\ref{sec:IIID}) as well as the growth-index parametrization
(Sec.~\ref{sec:IIIE}).
We then present a reconstruction of a model which has a weakened growth of structure relative to $\Lambda$CDM (Sec.~\ref{sec:IIIF}) before concluding with an example of a reconstruction from the inherently stable parameterization
introduced
in Sec.~\ref{sec:stableparam} (Sec.~\ref{sec:IIIG}).

The reconstructed Horndeski action is defined such that when
expanded up to second order in unitary gauge one recovers Eq.~\eqref{eftlag}
with the Horndeski functions given by~\cite{Kennedy:2017sof}
\begin{align}
G_{2}(\phi, X) = & -M_{*}^{2}U(\phi) - \frac{1}{2}M_{*}^{2} Z(\phi)X+a_{2}(\phi)X^{2} \nonumber\\
 &+\Delta G_{2} \,,
\label{eq:G2recon} \\
G_{3}(\phi,X) = & \: b_{0}(\phi)+b_{1}(\phi)X+\Delta G_{3} \,,
\label{eq:G3recon} \\
G_{4}(\phi, X) = & \: \frac{1}{2}M_{*}^{2}F(\phi)+c_{1}(\phi)X+\Delta G_{4} \,,
\label{eq:G4recon} \\
G_{5}(\phi, X)= & \: \Delta G_{5} \,.
\label{eq:G5recon}
\end{align}
Each term $U(\phi)$, $Z(\phi)$, $a_{2}(\phi)$, $b_{1}(\phi)$, $F(\phi)$, and $c_{1}(\phi)$ is expressed in terms of the EFT functions. See Table~I in the Appendix
for the full set of relations. Note that any contribution from $b_{0}(\phi)$ can be absorbed into $Z(\phi)$ after an integration by parts.
The $\Delta G_i$ terms are corrections one can add on to the action in Eqs.~\eqref{eq:G2recon}--\eqref{eq:G5recon} to move between Horndeski theories which are degenerate at the background and linear level. This reflects the nonlinear freedom in the family of reconstructed Horndeski models from linear theory.
It is worth noting that taking $c_T\simeq1$ as a linear constraint sets $c_1=0$ in Eq.~\eqref{eq:G4recon} but does not directly make a statement about $\Delta G_{4X/5}$.
However, excluding the highly unlikely cancellation of $c_1$ and $\Delta G_{4X/5}$, and assuming approximately linear theory from the outskirts of the Milky Way with $c_1=0$, the nonlinear contributions $\Delta G_{4/5}$ are still constrained by $|c_T-1|\lesssim10^{-13}$.

%%% FIGURE LCDM REF %%%
\begin{figure}%[b]
\resizebox{0.485\textwidth}{!}{
 \includegraphics{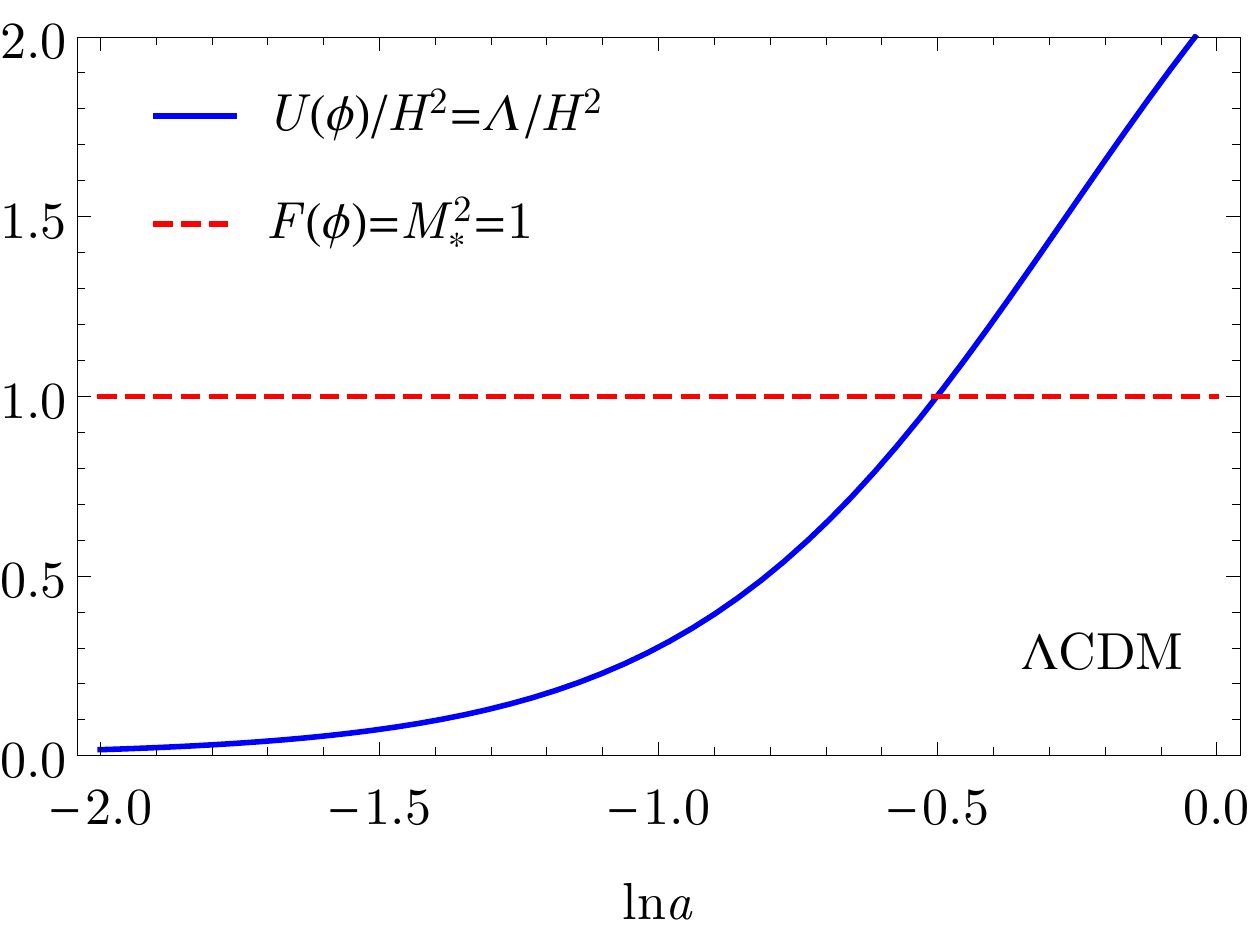}
 }
\caption{
Reconstructed contributions to the Horndeski action for $\Lambda$CDM, normalized with powers of $H$ received in the reconstruction (Table~I in
the Appendix).
The curves serve as reference for the comparison of the reconstructed modifications in Secs.~\ref{sec:IIIA}--\ref{sec:IIIG}.
Due to the normalization with $H^2$, the cosmological constant appears to decay at high redshift.}
\label{Refplot}
\end{figure}
%%% FIGURE LCDM REF %%%

In illustrations of the reconstructed
Horndeski functions
$G_i$, each
contributing
term is divided by the powers of $H$ it receives multiplying the EFT functions in the reconstruction (see Table~I in the Appendix).
This ensures a meaningful comparison of the effective modifications from $\Lambda$CDM rather than providing illustrations for deviations that are suppressed and do not propagate to the cosmological background evolution and linear perturbations.
For instance, we have $U(\phi)/H^{2}\sim b_{1}(\phi)/H\sim \bar{M}_{2}^{2}$.
As a reference, we show in Fig.~\ref{Refplot} the Horndeski functions $G_i$ that correspond to $\Lambda$CDM, where $G_{4}=1$, $G_{2}=-2\Lambda$ and $G_{3}=G_{5}=0$, i.e., in particular the term $\Lambda/H^2$.
The Planck 2015 value $\Omega_m=0.308$ \cite{Ade:2015xua} for the matter density parameter is adopted throughout the paper.
We also work in units where the bare Planck mass $M_{*}=1$,
such that the vertical axis on each plot indicates the deviation from this value.
Because the choice of how the scalar field is defined is arbitrary, we present the reconstructed terms as functions of $\ln a$ rather than $\phi$, except for the examples given in Secs.~\ref{sec:IIIA} and \ref{sec:IIIB}.
The colour scheme is set such that the terms in blue correspond to terms that can be identified in the matter sector, whereas the red terms couple to the metric and so in that sense are a ``modification'' of gravity. These modified gravity terms are $F(\phi)$ and $c_{1}(\phi)$, the latter being non-zero when the $\alpha_{T}=0$ constraint is dropped.

It is worth noting that one always has the freedom to redefine the scalar field $\phi$ in the action.
We shall briefly discuss how one can recast the reconstructed coefficients of the covariant theory
from functions of $\ln a$ to a more standard
description.
For this purpose, we choose the Brans-Dicke representation, where $F(\phi)\equiv \psi$, and then re-express all of the terms in the reconstructed action as a function of the new scalar field $\psi$.
This choice implies $\phi=F^{-1}(\psi)$ and
\begin{equation}
 \partial_{\mu}\phi =f(\psi)\partial_{\mu}\psi\, ,
\end{equation}
where for simplicity we have defined the function $f(\psi) \equiv d(F^{-1})/d\psi$.
After this field re-definition the reconstructed action written in terms of $\phi$ is transformed into
a scalar-tensor action for $\psi$ with
$(\partial \phi)^{2}=f^{2}(\psi)(\partial \psi)^{2}$ and $\Box \phi=f(\psi)\Box \psi+df/d\psi \, (\partial \psi)^{2}$. The new representation of the
theory then involves the terms
\begin{eqnarray}
 \tilde{U}(\psi) & = & U(\psi) \, , \\
 \tilde{Z}(\psi) & = & f^{2}(\psi)Z(\psi) \, , \\
 \tilde{b}_{1}(\psi) & = & f^{3}(\psi)b_{1}(\psi) \, , \\
 \tilde{a}_{2}(\psi) & = & a_{2}(\psi)f^{4}(\psi)+b_{1}(\psi)f^{2}(\psi)\frac{df}{d\psi} \, .
\end{eqnarray}
Depending on the functional form of $f(\psi)$ higher-derivative terms may be enhanced or suppressed in this description. For consistency, in this representation we also transform the Hubble parameter to be a function of $\psi$ such that $H\rightarrow \tilde{H}$.
We will show examples of this transformation in Secs.~\ref{sec:IIIA} and \ref{sec:IIIB}.

\subsection{Reconstruction of common EFT parameterizations}
\label{sec:IIIA}

%%% FIGURE COMMON PRAMS %%%
\begin{figure*}
\resizebox{0.495\textwidth}{!}{
\includegraphics{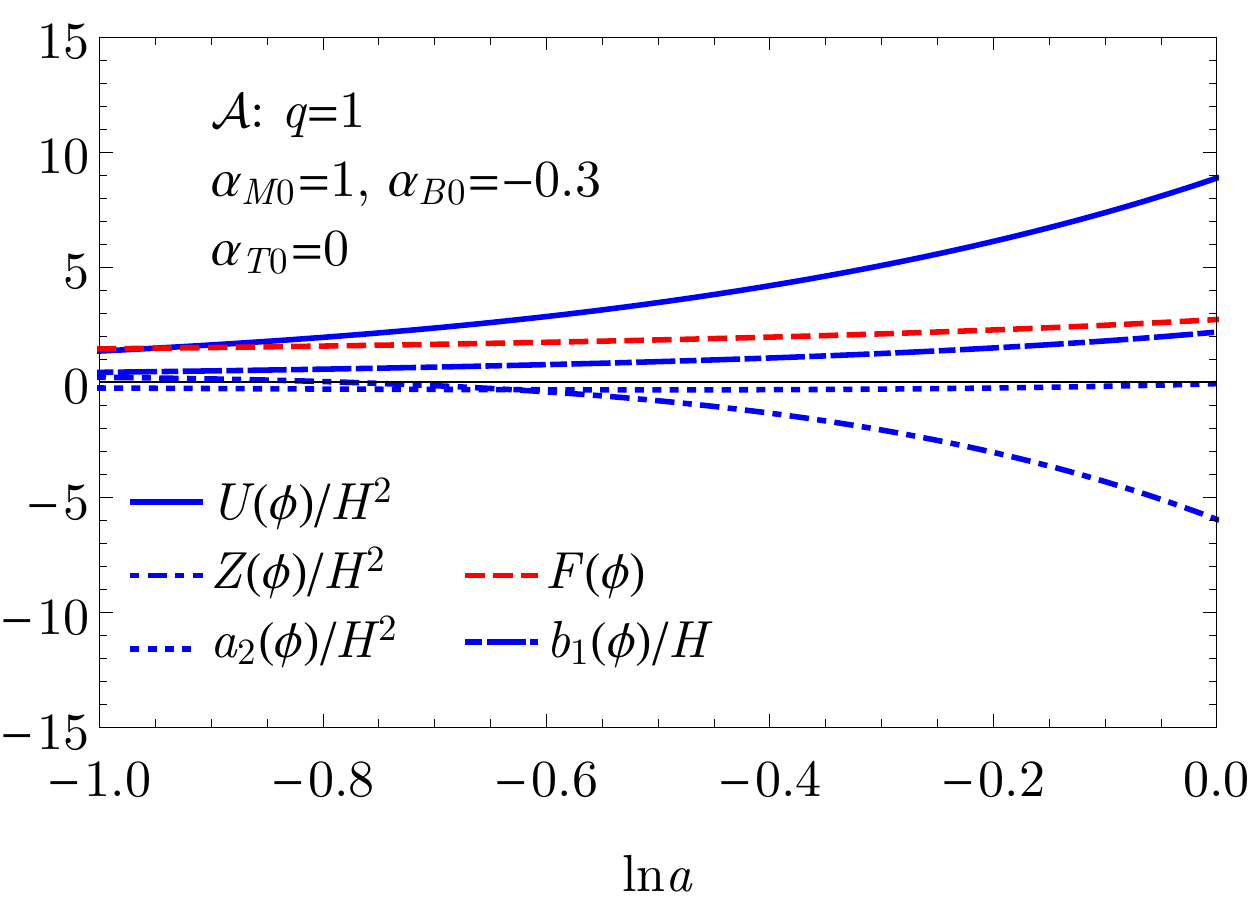}
}
\resizebox{0.495\textwidth}{!}{
\includegraphics{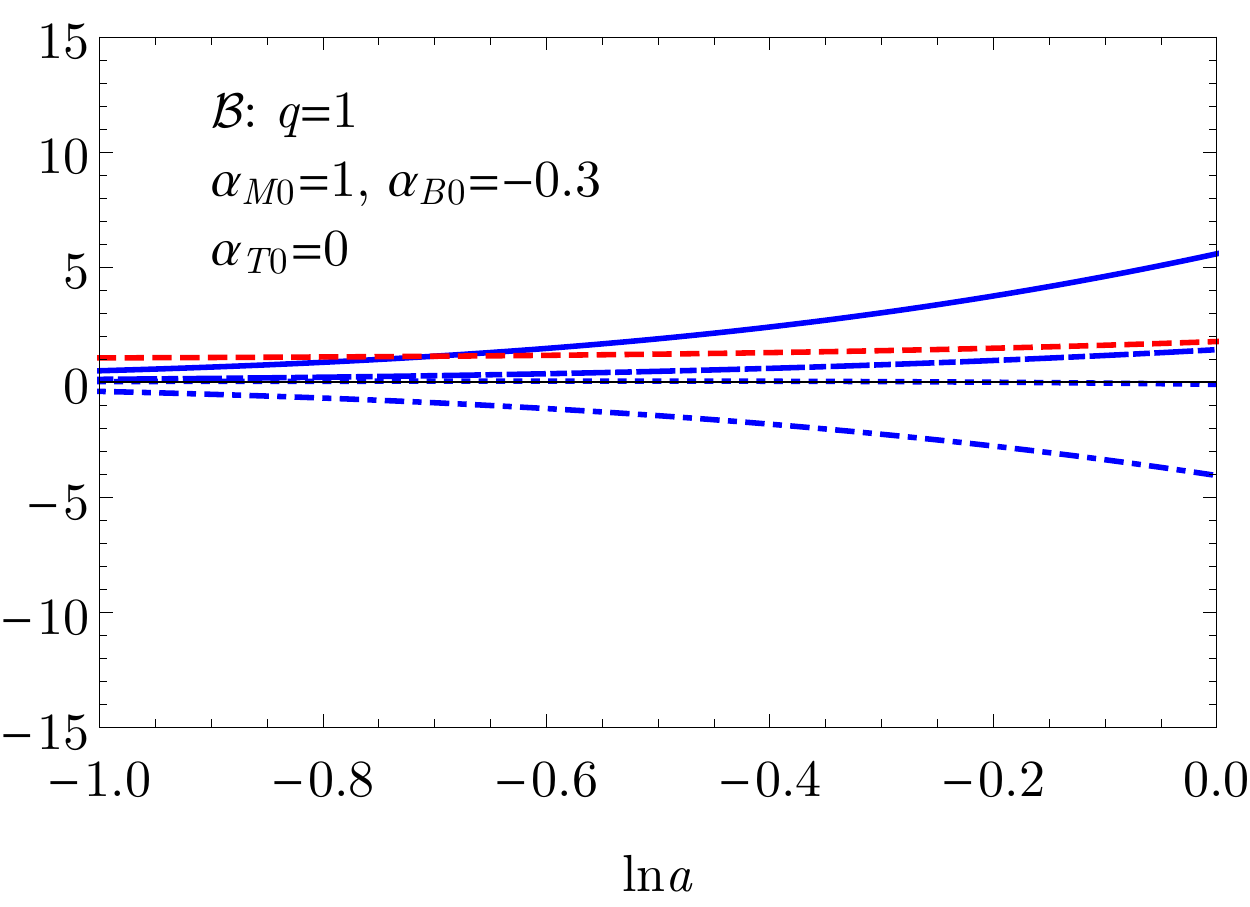}
}
\caption{Examples of reconstructed actions arising from two different parameterizations of the EFT functions $\mathcal{A}$ and $\mathcal{B}$ specified in Eqs.~\eqref{eq:par1} and \eqref{eq:par2}.
We chose equal amplitudes for the comparison.
The general evolution of the modifications is unaffected by the particular choice of time parametrization, although the magnitude of the various terms is enhanced when using parameterization $\mathcal{A}$. This can be attributed to the convergence to constant $\alpha_i$ at late times in $\mathcal{B}$.
The reconstructed terms of the scalar-tensor action can be converted into functions of a scalar field $\psi$, for instance, by adopting a Brans-Dicke representation and casting the functions in terms of $F\rightarrow\psi$ (see Fig.~\ref{BDRepplots}).
However, as the choice of scalar field is arbitrary, the reconstructions shall generally be illustrated as functions of $\ln a$.}
\label{Comparisonofparams}
\end{figure*}
%%% FIGURE COMMON PRAMS %%%

%%% FIGURE COMMON PRAMS BD %%%
\begin{figure*}
\resizebox{0.495\textwidth}{!}{
\includegraphics{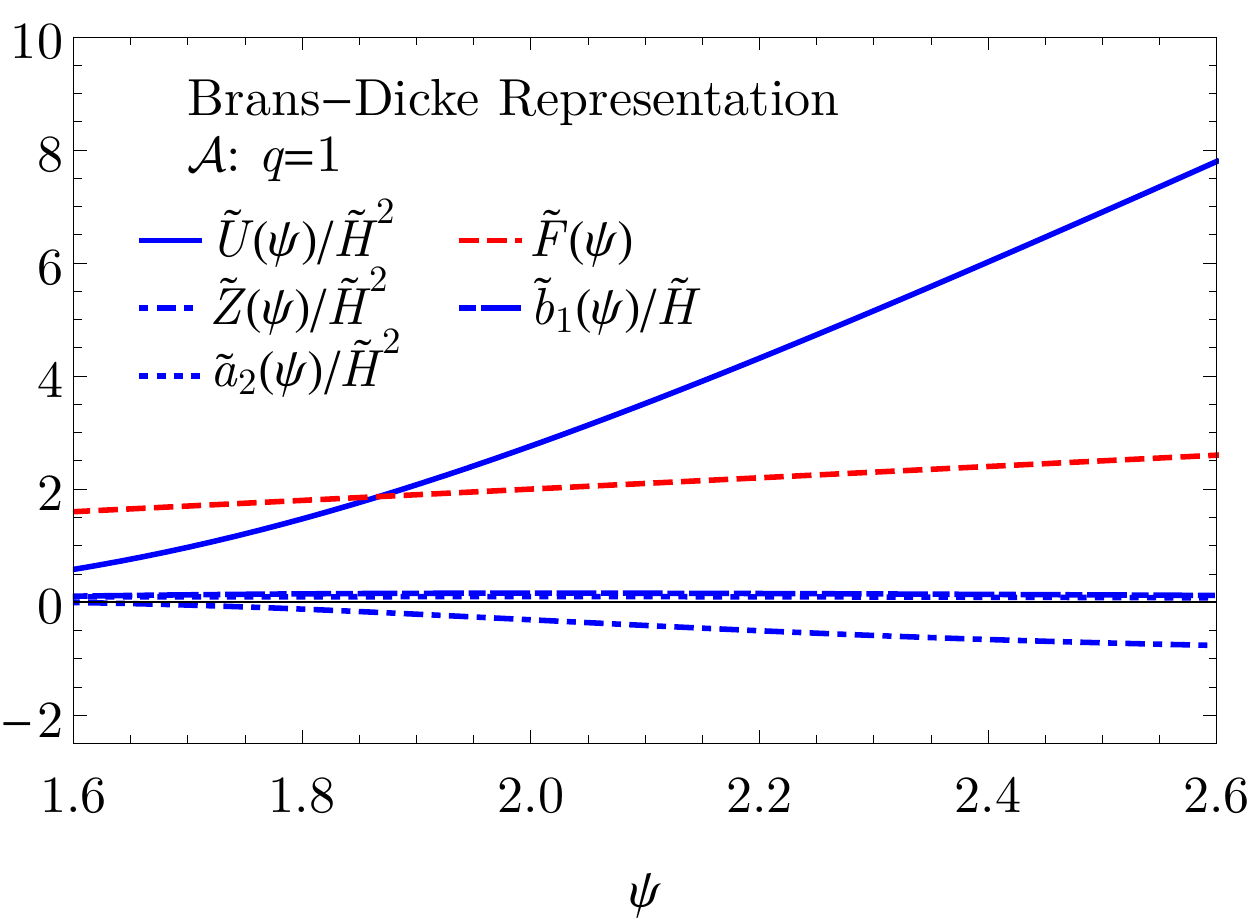}
}
\resizebox{0.495\textwidth}{!}{
\includegraphics{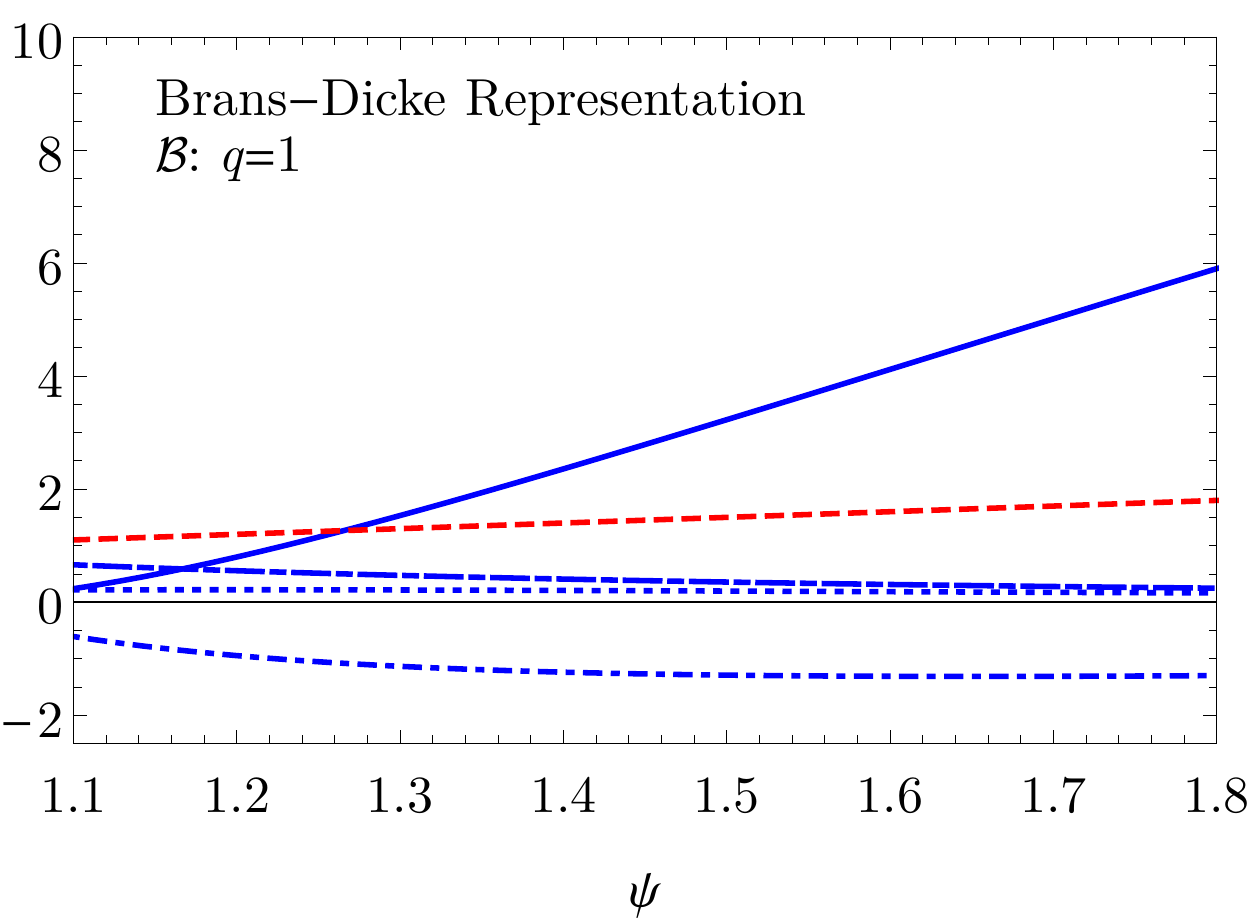}
}
\caption{Brans-Dicke representation, with $F(\phi) \equiv \psi$, of the reconstructed scalar-tensor theories illustrated in Fig.~\ref{Comparisonofparams}. %We have kept each $\alpha_{i0}$ the same as in Fig.~\ref{Comparisonofparams}
%\llo{What do you mean? This is not the real amplitude? Or do you mean the domain of $\psi$ corresponds to the range of $\ln a$?}
%so that this figure can be directly compared with Fig.~\ref{Comparisonofparams}. \jk{Is this clearer?}
%\llo{I see. I don't think we need to mention this explicitly as it seems clear from the context that it should be the same model with the same effective modifications.}
We have transformed the Hubble rate $H\rightarrow\tilde{H}$ such that it is also a function of $\psi$ and divided each term in the action by appropriate powers of $\tilde{H}$ (see Sec.~\ref{sec:results}).
}
\label{BDRepplots}
\end{figure*}
%%% FIGURE COMMON PRAMS BD %%%

A common choice of phenomenologically 
motivated
functional forms of the EFT modifications
is to parameterize them
in such a way that they only become relevant at late times. Typically their evolution is tied to the scale factor $a(t)$ or to the dark energy density $\Omega_{\Lambda}(a)\equiv H_{0}^{2}\Omega_{\Lambda}/H^{2}$ raised to some power $q$.
Note that now, with the GW170817 constraint, self-acceleration from modified gravity is
strongly challenged
as a direct explanation for the late-time accelerated expansion~\cite{Lombriser:2016yzn} and it can be questioned whether the functional form of such parameterizations continues to be well motivated.
On the other hand, a dark energy model may still introduce a related modification of gravity, for instance, as a means to remedy the old cosmological constant problem of a non-gravitating vacuum.
We set this issue aside for now
and adopt the two parametrizations
\begin{eqnarray}
 \mathcal{A} & : &  \alpha_{i}=\alpha_{i0} a(t)^{q_{i}} \,, 
\label{eq:par1} \\
 \mathcal{B} & : & \alpha_{i}=\alpha_{i0}\left(\frac{\Omega_{\Lambda}(a)}{\Omega_{\Lambda,0}} \right)^{q_{i}} \,.
\label{eq:par2}
\end{eqnarray}
Here the label $i$ runs over the set of functions $\left\{i \in M,T,K,B \right\}$
in Eqs.~\eqref{Am}--\eqref{At}.
The two parametrizations can be used to study the effect of small deviations from $\Lambda$CDM in the linear late-time fluctuations
resulting from a set of non-vanishing $\alpha_i$. 

In principle, there are many alternative parameterizations that could be used beyond these simple ones.
For the purposes of this paper we shall however restrict ourselves to 
these two examples 
which have been frequently used
in the literature (see e.g.~Ref.~\cite{Ade:2015rim}).
It was recently suggested that
those are 
sufficiently general to encompass the linear effects of the different time dependencies in a variety of modified gravity theories~\cite{Gleyzes:2017kpi}
(however, also see Ref.~\cite{Linder:2015rcz}).
The reconstruction from EFT back to manifestly covariant theories provides a method to examine how the underlying covariant theory changes with a different choice of parameterization. One can thus begin to address the question of what scalar-tensor theory is actually being constrained when a particular parameterization is adopted.

To provide concrete examples for the
models that are reconstructed from
Eqs.~\eqref{eq:par1} and \eqref{eq:par2},
we parametrize $\alpha_{M}$, $\alpha_{B}$ and $\alpha_{T}$ with $\mathcal{A}$ or $\mathcal{B}$ and then set $\alpha_{K}$ such that
$c_{s}^{2}=1$ (see discussion in Sec.~\ref{sec:IIA}).
Note that strictly speaking this deviates from adopting Eqs.~\eqref{eq:par1} and (\ref{eq:par2}) for all $\alpha_i$ but it simplifies the stability treatment of the model. Furthermore, the deviation is only relevant on near-horizon scales.
Numerical values for $\alpha_{i0}$ are then chosen to ensure that the stability condition $\alpha>0$ in Eq.~\eqref{Stabcriteria1} is satisfied.
For illustration, we set $\alpha_{M0}=1$, $\alpha_{B0}=-0.3$ and $\alpha_{T0}=0$ with $q_i=q=1$.
This yields a stable scalar-tensor theory for both parameterizations $\mathcal{A}$ and $\mathcal{B}$.
The corresponding terms of the Horndeski functions are shown in Fig~\ref{Comparisonofparams}.
The behavior of the reconstructed theories is tied to the functional form of the parameterization used, with the Horndeski modifications becoming more relevant at later times.
We note that the general form of these modifications is independent of the particular parametrization adopted between $\mathcal{A}$ and $\mathcal{B}$.
However,
one can identify minor differences.
For instance, the magnitude of the reconstructed modifications for $\mathcal{A}$ are larger.
This is due to saturation of the modifications in $\mathcal{B}$ at late times.
This particular choice for each $\alpha_{i0}$ leads to a model with an enhanced potential relative to $\Lambda$CDM and the standard kinetic term $Z(\phi)$ dominating the action at late times. There is a small contribution from the cubic term $b_{1}(\phi)$ but the k-essence term $a_{2}(\phi)$ is negligible.  
In the Appendix we present a number of examples which examine the sensitivity of the reconstruction to changes in $\alpha_{i0}$ and $q_{i}$.
For instance, by increasing the powers $q_i$ in each of the parameterizations, the effects of modified gravity and dark energy
are delayed to later times.
Changing the amplitude of each $\alpha_{i0}$ on the contrary has a
larger effect on the underlying theory. For example, when $\alpha_{B}$ dominates over $\alpha_{M}$ the cubic Galileon term $b_{1}(\phi)$ dominates over the potential $U(\phi)$ at late times, whereas when $\alpha_{M}$ dominates over $\alpha_{B}$ the potential and kinetic term $Z(\phi)$ are enhanced with smaller contributions from the k-essence and cubic Galileon terms. 
However, we find that the mapping is relatively robust, with small deviations in the $\alpha_i$ parameters around some fixed values not significantly altering the underlying theory.
While we have checked this for a number of examples, further work is necessary to investigate this aspect more thoroughly.
More details can be found in the Appendix.

Finally, in Fig.~\ref{BDRepplots} we illustrate the corresponding Brans-Dicke representations of the reconstructed theories for $\mathcal{A}$ and $\mathcal{B}$ that are presented in Fig.~\ref{Comparisonofparams}.
In this
description
the behavior of each term in the reconstruction is now dependent on the evolution of $F(\phi)$. 
It is clear that
the functional form of each term in the theory remains broadly similar whether parameterization $\mathcal{A}$ or $\mathcal{B}$ is chosen.

%%%%%%%%%%%%%%%%%%%%%%%  MIN S.A  %%%%%%%%%%%%%%%%%%%%%%%%%%%%%%%%%%%%%5

\subsection{Minimal self-acceleration}
\label{sec:IIIB}

%%% FIGURE MINIMAL SELF-ACC %%%
\begin{figure*}
\resizebox{0.496\textwidth}{!}{
\includegraphics{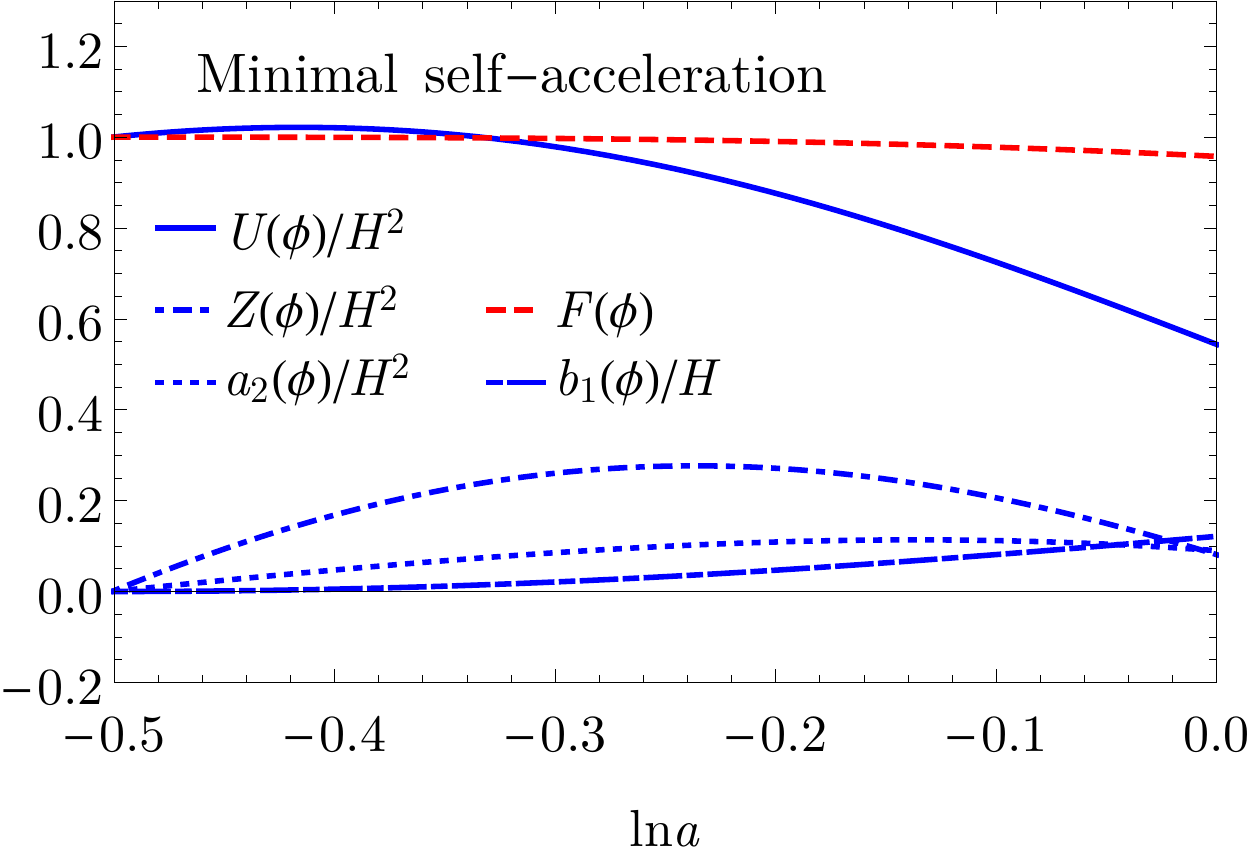}
}
\resizebox{0.495\textwidth}{!}{
\includegraphics{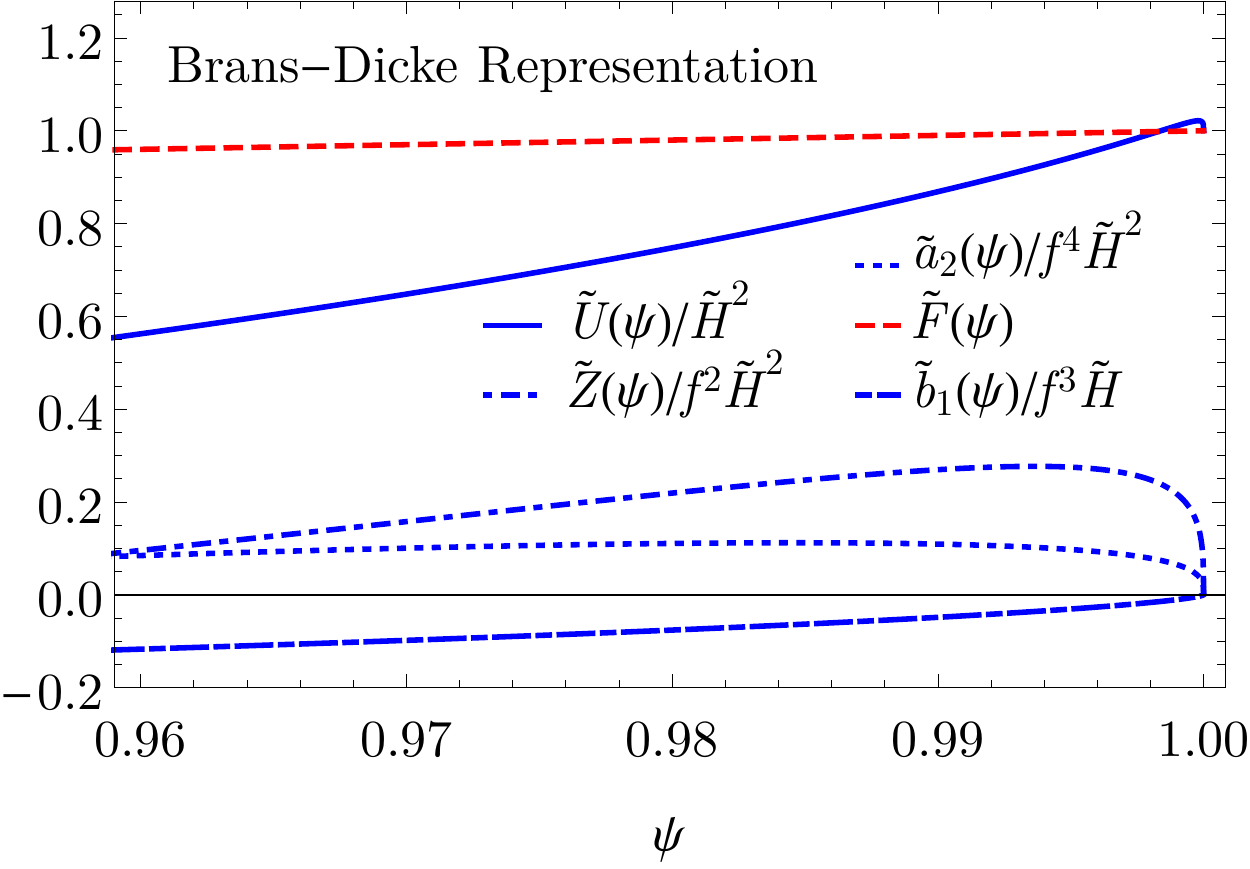}
}
\caption{\emph{Left}: The scalar-tensor theory yielding the minimal modification of gravity required for self-acceleration with $c_T=1$.
Note that the scalar field potential at early times ensures a recovery of the decelerating phase of $\Lambda$CDM and decays in the accelerating phase $H^2<\Lambda$ to barely prevent positive acceleration in
Einstein frame.
\emph{Right}: The minimal self-acceleration model expressed using the Brans-Dicke representation in terms of $\psi$. We have divided each term by the corresponding factors of $f(\psi)$ for a clearer comparison to the left-hand panel.
Note that as $F(\phi)$ is decreasing, the forward direction in time corresponds to
decreasing values of $\psi$.
}
\label{MinSAPlots}
\end{figure*}
%%% FIGURE MINIMAL SELF-ACC %%%

The LIGO/Virgo constraint of $|c_T-1|\lesssim\mathcal{O}(10^{-15})$ and its implication that a genuinely self-accelerated Universe in scalar-tensor gravity must
be attributed to a significant evolution in $M^2$ was first anticipated in Ref.~\cite{Lombriser:2015sxa}. This trivially excludes acceleration arising from an evolving speed of gravity $c_T$ and the according class of gravitational models such as genuinely self-accelerated quartic and quintic Galileons
and their Horndeski
and higher-order generalizations with $\alpha_T\neq0$,
i.e., $G_{4X}, G_5\neq0$ for Horndeski gravity (see e.g.~Ref.~\cite{Kimura:2011qn}).
With this expectation, Ref.~\cite{Lombriser:2016yzn} devised the minimal surviving modification of gravity that can yield cosmic self-acceleration consistent with an event like GW170817.
We briefly review this model, before presenting a corresponding reconstructed covariant scalar-tensor theory.

While self-acceleration may generally be defined as cosmic acceleration without a cosmological constant or a scalar field potential, this definition includes exotic dark energy models like k-essence \cite{ArmendarizPicon:1999rj} or cubic Galileon and Kinetic Gravity Braiding (KGB) \cite{Deffayet:2010qz} models.
Hence, a
more precise definition is required if cosmic acceleration is genuinely to be attributed to an intrinsic modification of gravity.
This definition also needs to distinguish between models where dark energy or a cosmological constant drives cosmic acceleration but where a modification of gravity may still be present.
As a definition of a genuinely self-accelerated modification of gravity in chameleon gravity models, Ref.~\cite{Wang:2012kj} argued that while cosmic acceleration should be present in the Jordan frame with metric $g_{\mu\nu}$, it should not occur in the conformally transformed Einstein frame $\tilde{g}_{\mu\nu}=\Omega g_{\mu\nu}$ with the conformal factor $\Omega$.
Otherwise, the acceleration should be attributed to an exotic matter contribution.
In Ref.~\cite{Lombriser:2015sxa} this argument was generalized to include an evolving speed of gravity $c_T$ in addition to an evolving strength of gravity $M^{-2}$ as the cause of self-acceleration. This encompasses the quartic and quintic Galileon models as well as their generalizations in the full Horndeski action and beyond.
These effects can be described by an effective conformal factor in the cosmological background that absorbs the contributions from conformal and disformal couplings in the Einstein frame.
An Einstein-Friedmann frame can then be defined from the effective conformal (or pseudo-conformal) transformation of the cosmological background.
Alternatively, this can be viewed as assigning genuine cosmic self-acceleration to the magnitude of the breaking of the strong equivalence principle \cite{Joyce:2016vqv}.
Note that self-acceleration arising from a dark sector interaction would correspondingly be attributed to the breaking of the weak equivalence principle.

With this definition, genuine self-acceleration implies that in the Einstein-Friedmann frame
\begin{equation}
\frac{d^{2}\tilde{a}}{d\tilde{t}^{2}} \leq 0  \, ,
\end{equation}
with the minimal modification obtained at equality.
From inspection of the transformed Friedmann equations, it follows that this condition can hold only if the EFT function $\Omega$ satisfies~\cite{Lombriser:2015sxa}
\begin{equation}
 -\frac{d \ln \Omega}{d \ln a} \gtrsim \mathcal{O}(1) \, .
\label{SAcondition}
\end{equation}
Note that
\begin{equation}
 \Omega=\frac{M^{2}}{M_{*}^{2}}c_{T}^2 \,,
 \label{consistencyrelation}
\end{equation}
implying that self-acceleration requires a significant deviation in the speed of gravitational waves or an evolving Planck mass.
Since GW170817 strongly constrains the deviations of $c_T$ at low redshifts, i.e., in the same regime of cosmic acceleration,
one can set $c_T=1$ ($\alpha_T=0$) in Eq.~\eqref{consistencyrelation}, so that self-acceleration must solely arise from the effect of $M^2$ (or $\alpha_M$)~\cite{Lombriser:2015sxa}.
The minimal modification of gravity for genuine cosmic self-acceleration can then be derived by minimizing the impact of a running $M^2$ on the large-scale structure.
For Horndeski gravity, this implies $\alpha_B=\alpha_M$ with $c_{s}^{2}=1$ setting $\alpha_K$~\cite{Lombriser:2016yzn}.
The EFT functions of the model are then fully specified by a given expansion history $H(z)$, which for a minimal departure from standard cosmology can be set to match $\Lambda$CDM.
We present the reconstructed scalar-tensor action for minimal genuine self-acceleration in Fig.~\ref{MinSAPlots}.

Note that for a $\Lambda$CDM expansion history, cosmic acceleration in Jordan frame occurs when $H^2 < \Lambda$.
Hence, a \emph{minimal} self-acceleration must recover $U/H^2=1$ at the transition from a decelerating to an accelerating cosmos.
There is therefore still a scalar field potential or cosmological constant that contributes to reproduce the $\Lambda$CDM expansion history in the decelerating phase where there are no modifications of gravity but then it decays at a rate so as not to introduce any positive acceleration in the Einstein-Friedmann frame, keeping the Universe at a constant expansion velocity.
The cosmic acceleration in Jordan frame is then solely driven by the decaying Planck mass, commencing at the threshold $H^2 < \Lambda$.
It is in this sense a model with the minimal gravitational modification required for positive acceleration.
Alternatively, the scalar field potential could be removed by hand, but this would lead to a loss of generality and the conservative character of the inferred conclusions.

The reconstructed scalar-tensor terms $F(\phi)$ and $U(\phi)$ for minimal self-acceleration in Fig.~\ref{MinSAPlots} are decaying functions as expected, with the behavior of the other terms acting to minimize the impact on scalar perturbations and
the large-scale structure.
At redshift $z=0$, we find comparable contributions from the quintessence $Z(\phi)$, k-essence $a_{2}(\phi)$ and cubic Galileon $b_{1}(\phi)$ terms indicating that they are all required to ensure a minimal self-acceleration.
Ref.~\cite{Lombriser:2016yzn} performed a MCMC analysis of the model with recent cosmological data, finding a $3\sigma$ worse fit than $\Lambda$CDM and hence strong evidence for a cosmological constant over the minimal modification of gravity required in Horndeski scalar-tensor theories for self-acceleration and consistent with the expectation of the GW170817 result.
The constraints are driven by the cross correlation of the integrated Sachs-Wolfe effect with foreground galaxies.
It is worth noting that the minimal self-acceleration derived for $M^2$ also applies to beyond-Horndeski \cite{Zumalacarregui:2013pma, Gleyzes:2014dya} theories or Degenerate Higher-Order Scalar-Tensor (DHOST) theories \cite{Langlois:2015cwa}.
Due to the additional free EFT functions introduced in those models, however, the measurement of $\alpha_T\simeq0$ is not sufficient to break the dark degeneracy and linear shielding is still feasible~\cite{Lombriser:2014ira}.
However, it was pointed out in Ref.~\cite{Lombriser:2015sxa} that Standard Sirens tests of the evolution of $M^2$ are not affected by this degeneracy and may provide a $5\sigma$ result on minimal self-acceleration for Horndeksi gravity and its generalizations over the next decade.
Independently of future gravitational wave measurements, minimal self-acceleration provides a benchmark model which can quantify to what extent galaxy-redshift surveys like Euclid \cite{Laureijs:2011gra, Amendola:2012ys} or LSST \cite{Ivezic:2008fe} can exclude cosmic self-acceleration from modified gravity, precluding dark degeneracies (or linear shielding) in higher-order gravity.

%%%%%%%%%%%%%%%%%% Linear Shielding %%%%%%%%%%%%%%%%%%%%%%%%%%%%%%%%
\subsection{Covariant model with Linear Shielding}
\label{sec:IIIC}

%%% FIGURE LIN SHIELD %%%
\begin{figure}
\resizebox{0.485\textwidth}{!}{
\includegraphics{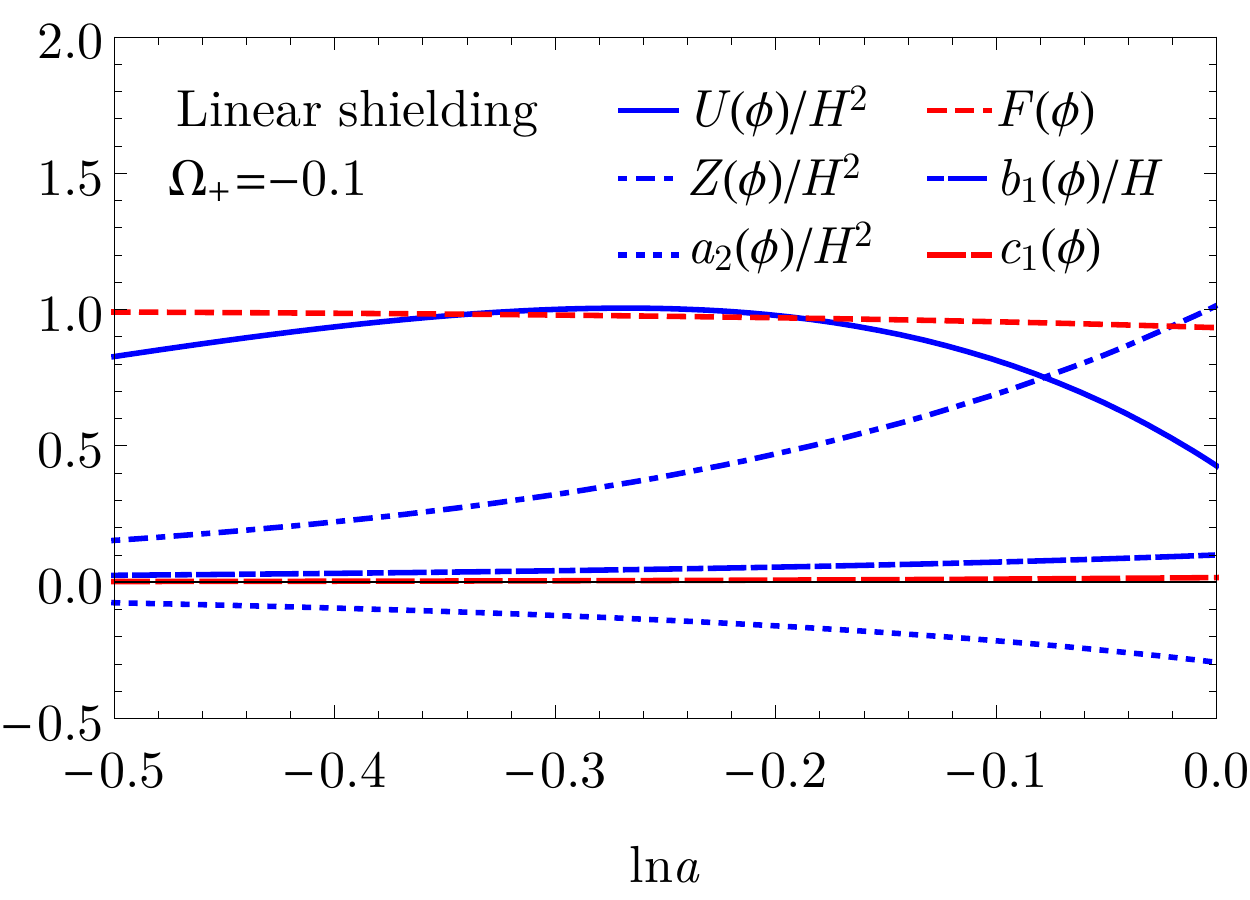}
}
\caption{The scalar-tensor theory that exhibits linear shielding for the parameterization in Eq.~\eqref{OmegaParam}.}
\label{LinearShielding}
\end{figure}
%%% FIGURE LIN SHIELD %%%

A number of classes of scalar-tensor theories that cannot be distinguished from concordance cosmology via observations of the large-scale structure and background evolution alone were presented in Ref.~\cite{Lombriser:2014ira}.
This phenomena arises through a linear shielding mechanism.
It was then shown in Ref.~\cite{Lombriser:2015sxa} that for Horndeski theories the measurement of $\alpha_{T} = 0$ breaks this degeneracy.
However, linear shielding still remains viable in more general scalar-tensor theories and its extension to the modified gravitational wave propagation may even provide a means to evade the GW1701817 constraint for self-acceleration from $c_T$ \cite{Battye:2018ssx}.
It is furthermore worth considering that the $\alpha_{T}\simeq0$ constraint only applies at late times and it may remain of interest to examine Horndeski models with non-vanishing $\alpha_{T}$ at higher redshifts that may also undergo linear shielding.
It is therefore worthwhile to examine some basic forms of the scalar-tensor theories that give rise to linear shielding.

In order to recover $\Lambda$CDM in the linear cosmological small-scale limit,
for models belonging to the $\mathcal{M}_{\textnormal{II}}$ class of linear shielding, 
the EFT functions must satisfy the conditions~\cite{Lombriser:2014ira,Lombriser:2015sxa} 
\begin{align}
 \alpha_{M}M^{2} & =  \alpha_{B}\kappa^{2}M^{4} -\frac{1-\kappa^{2}M^{2}}{\alpha_{B}} \nonumber \label{LS1}\\
  & \times \left\{\frac{\rho_{m}}{2H^{2}}+\left[ \alpha_{B}^{\prime}+\alpha_{B}+(1+\alpha_{B})\frac{H^{\prime}}{H}   \right]M^{2}   \right\} \, , \\
 \alpha_{T} & =  \frac{\kappa^{2}M^{2}-1}{(1+\alpha_{B})\kappa^{2}M^{2}-1} \alpha_{M} \, .
 \label{LS2}
\end{align}
Applying these constraints, setting the background expansion to match $\Lambda$CDM and fixing $c_{s}^{2}=1$ leaves one free EFT function.
With a parameterization of this function and applying our reconstruction, one can then find a scalar-tensor theory that exhibits linear shielding. 

Here we adopt the same parameterization as Ref.~\cite{Lombriser:2014ira} and choose
\begin{equation}
\Omega(a)=1+\Omega_{+}a^{n} \, ,
\label{OmegaParam}
\end{equation}
with $\Omega_{+}=-0.1$ and $n=4$.
The general behavior of all the terms in the reconstruction of this linear shielding model is fairly insensitive to changing the magnitude of $\Omega_{+}$, the one free parameter in the model.
The action does differ under a change in the sign of $\Omega_{+}$, but this acts to decelerate the expansion.

We illustrate the reconstructed scalar-tensor action for our choice of parameters in Fig.~\ref{LinearShielding}.
$U(\phi)$ is dominated by the EFT function $\Lambda(t)$ which behaves in a similar way to the minimal self-acceleration model, acting as a cosmological constant at early times before decaying away at late times.
The late-time decay of $\Lambda(t)$ is compensated by the other terms in the reconstruction to ensure that the linear perturbations are not affected in their $\Lambda$CDM behavior.
$F(\phi)$ also decays which is a consequence of our choice of a negative $\Omega_{+}$, required for self-acceleration. 
The linearly shielded Horndeski model requires a decrease in the speed of gravitational waves over time which leads to  $c_{1}(\phi)$ growing in time. The kinetic terms become more dominant at late times, predominantly being driven by $\Gamma$ and $M_{2}^{4}$ with the form of $a_{2}(\phi)$ essentially mimicking that of $M_{2}^{4}$. 
In $b_{1}(\phi)$ the contributions $\bar{M}_{1}^{3}$ and $\bar{M}_{2}^{2}$ compete
and suppress it relative to the other terms in the action.

Although the conditions for linear shielding may seem contrived when expressed in terms of the EFT parameters, we find that it is nevertheless the case that there is a 
generic scalar-tensor theory which gives rise to this mechanism for the particular parameterization we adopt.
It is also worth bearing in mind that observational large-scale structure constraints allow for a broad variation around the strict conditions in Eqs.~\eqref{LS1} and \eqref{LS2} in which the model space remains observationally degenerate with $\Lambda$CDM.
%

%%%%%%%%%%%%%%%%%%%%%%%%%%%%%%%%%% Mu and Gamma reconstruction %%%%%%%%%%%%%%%%%%%%%%%%%%%%%%%%%%%%5

\subsection{$\mu$ and $\eta$ reconstruction}
\label{sec:IIID}
%

%%% FIGURE MU GAMMA %%%
\begin{figure*}
\resizebox{0.492\textwidth}{!}{
\includegraphics{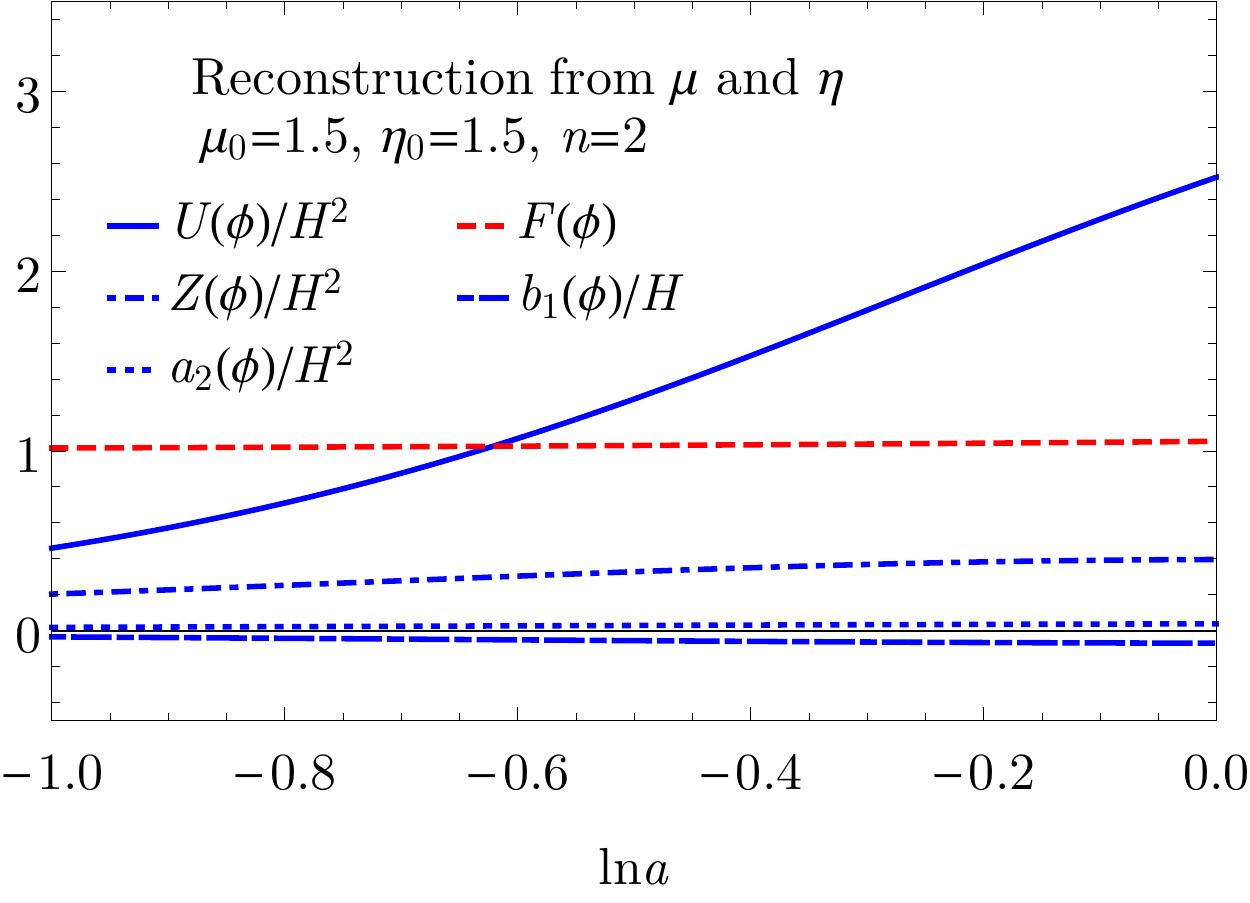}
}
\resizebox{0.497\textwidth}{!}{
\includegraphics{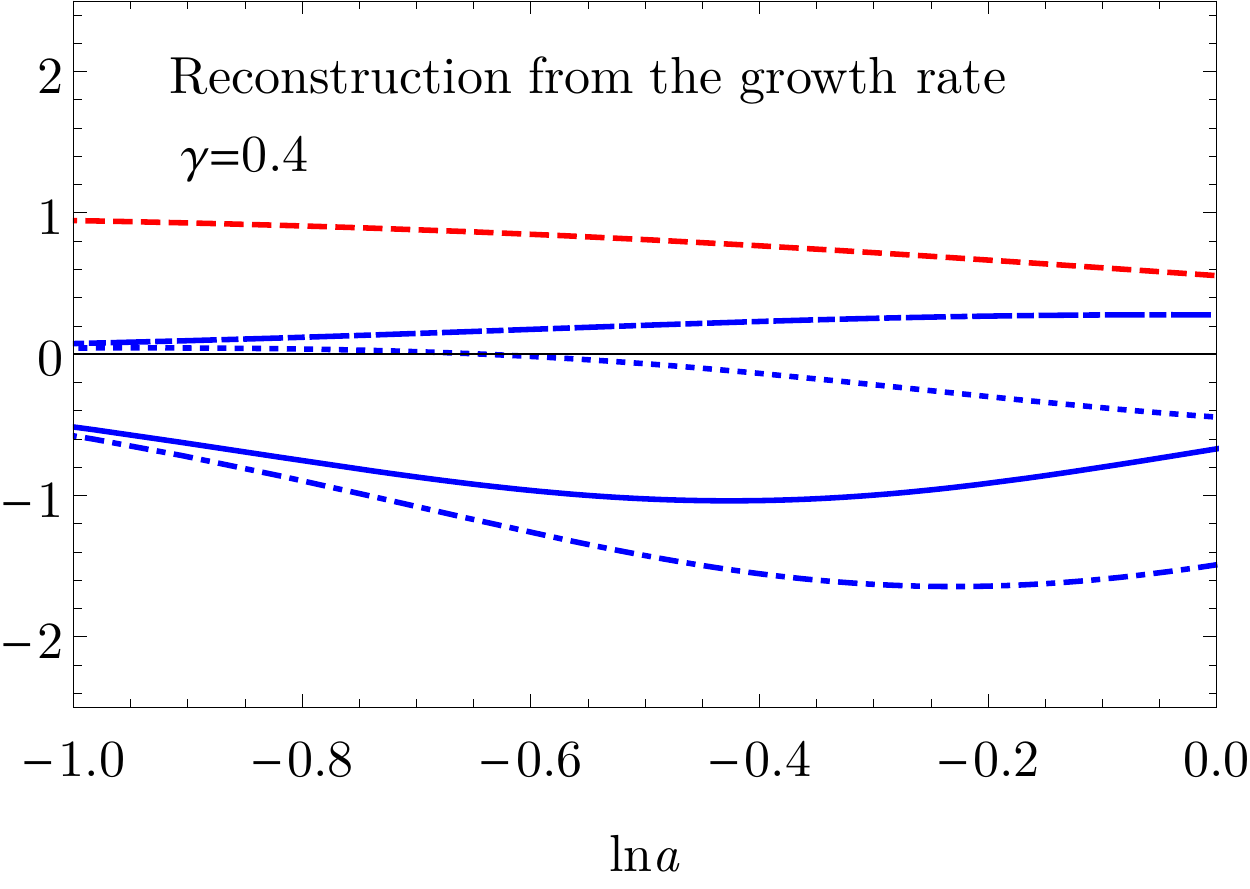}
}
\label{Muandgammazoomed}\label{Growthrate}
\caption{\emph{Left}: Reconstructed action from a direct parameterization of the modified Poisson equation and the gravitational slip. \emph{Right}: Reconstructed action from the growth-index parametrization.
%rate trigger parameter $\gamma$.
}
\label{Growthrateandmuandgamma}
\end{figure*}
%%% FIGURE MU GAMMA %%%

The effects of modified gravity and dark energy on the large-scale structure can be described phenomenologically by the behavior of two functions of time and scale that parameterize a deviation in the Poisson equation $\mu(a,k)$ and introduce a gravitational slip $\eta(a,k)$~\cite{Uzan:2006mf, Amendola:2007rr, Caldwell:2007cw, Hu:2007pj, Zhang:2007nk}.
We shall work with a perturbed FLRW metric in the Newtonian gauge with $\Psi \equiv \delta g_{00}/2g_{00}$ and $\Phi\equiv\delta g_{ii}/2g_{ii}$
and matter density perturbations $\Delta_{m}$ in the comoving gauge.
The effects of modified gravity and dark energy on the perturbations can be described via the relations 
\begin{equation}
k_{H}^{2}\Psi =-\frac{\kappa^{2}\rho_{m}}{2H^{2}}\mu(a,k)\Delta_{m} \, ,  
\label{Poisson}
\end{equation}
\begin{equation}
\Phi=-\eta(a,k) \Psi \, ,
\label{slip}
\end{equation}
where $k_{H}\equiv k/(aH)$.
Energy and momentum conservation then closes the system of differential equations and one can solve for the evolution of the linear perturbations.

The modifications $\mu(a,k)$ and $\eta(a,k)$ are more general than the EFT formalism but the two can be linked in the domain covered by the EFT functions.
Specifically, in the formal linear theory limit of $k\rightarrow \infty$ the functions $\mu$ and $\eta$ can be treated as only functions of time. In this limit, they can be related to the EFT functions via
\begin{align}
\mu_{\infty} & =  \frac{2\left[\alpha_{B}(1+\alpha_{T})-\alpha_{M}+\alpha_{T}   \right]^{2}+\alpha(1+\alpha_{T})c_{s}^{2}}{\alpha c_{s}^{2} \kappa^{2}M^{2}} \, ,
\label{muinf} \\
\eta_{\infty} & =  \frac{2\alpha_{B}\left[\alpha_{B}(1+\alpha_{T})-\alpha_{M}+\alpha_{T}\right]+\alpha c_{s}^{2}}{2\left[\alpha_{B}(1+\alpha_{T})-\alpha_{M}+\alpha_{T} \right]^{2}+\alpha(1+\alpha_{T})c_{s}^{2}} \, .
\label{gammainf}
\end{align}

For the purposes of this paper we shall remain in
this small-scale
regime
and parameterize the time-dependent modifications as
\begin{equation}
\mu(a)=1+(\mu_{0}-1) a^{n} \, ,
\end{equation}
\begin{equation}
\eta(a)=1+(\eta_{0}-1)a^{n} \, ,
\end{equation}
with $n=2$.
For simplicity, we furthermore consider a background evolution $H(t)$ that matches that of $\Lambda$CDM
and we adopt $\alpha_{T}=0$ at all times to break the degeneracy in parameter space.
The kineticity function $\alpha_{K}$ is set by the choice
$c_{s}^{2}=1$.
The set of EFT functions is then closed by Eqs.~\eqref{muinf} and \eqref{gammainf}, determining the evolution of $\alpha_{B}$
and $\alpha_{M}$.
Given a choice of parameters $\mu_0$, $\eta_0$ we can now
reconstruct a corresponding Horndeski scalar-tensor theory.
For this example we choose a model that exhibits both a non-zero gravitational slip and an enhanced growth of structure today by setting $\mu_{0}=\eta_{0}=3/2$.

The reconstructed scalar-tensor action is illustrated in Fig.~\ref{Growthrateandmuandgamma}.
The dominant term at redshift zero is $U(\phi)$. It behaves as a cosmological constant which is enhanced relative to its $\Lambda$CDM value. $F(\phi)$ is determined through the evolution of $M^{2}$. Despite the enhanced growth with this parameterization of $\mu$ and $\eta$ the Planck mass increases from its GR value today. The enhanced growth is therefore coming from the clustering effect of $\alpha_{B}$. This can be seen more clearly by writing
\begin{equation}
\mu=\frac{M_{*}^{2}}{M^{2}}\left(1+\frac{2(\alpha_{B}-\alpha_{M})^{2}}{\alpha c_{s}^{2}} \right) \, .
\end{equation}
Although the Planck mass is increasing, $\alpha_{B}$ also increases to dominate over $\alpha_{M}$ and gives rise
to the pre-defined
evolution in $\mu(a)$.
The domination of $\alpha_{B}$ over $\alpha_{M}$ also causes $b_{1}(\phi)$ to be negative.
This is because $b_{1} \sim \bar{M}_{1}^{3} \sim (\alpha_{M}-2\alpha_{B})$ up to numerical factors and positive background terms. In this model $\alpha_{K}\approx 0$. The background terms that contribute to $M_{2}^{4}$ compete to cancel each other out. The dominant term in $a_{2}(\phi)$ is from $-\bar{M}_{1}^{3}$ or $\alpha_{B}$, which is small and positive.

%%%%%%%%%%%% Reconstruction from the growth rate %%%%%%%%%%%%%%%%%%%%%

\subsection{$\Omega_{m}^{\gamma}$ reconstruction}
\label{sec:IIIE}

One of the most commonly used formalisms for testing departures from GR with the large-scale structure is the growth-index parametrization~\cite{PeeblesLSS, Wang:1998gt, Linder:2005in}.
It involves a direct parameterization of a modification of the growth rate
\begin{equation}
f \equiv \frac{d \, \textnormal{ln} \, \Delta_{m}(a,k)}{d \, \textnormal{ln} \, a} =\Omega_{m}(a)^{\gamma}
\label{Triggerrelation}
\end{equation}
with the growth-index parameter $\gamma$,
which is generally considered a \emph{trigger} or \emph{consistency} parameter.
Any observational deviation from its
GR value $\gamma\approx6/11$ \cite{PeeblesLSS} will indicate a breakdown of GR.

On sub-horizon scales ($k\gg aH$) the modified growth equation for the matter density contrast is given by
\begin{equation}
\Delta_m''+\left(2+\frac{H^{\prime}}{H} \right)\Delta_m'-\frac{3}{2}\Omega_{m}(a)\mu_{\infty}(a)\Delta_m=0 \,,
\label{growthequation}
\end{equation}
which follows from the modified Poisson equation \eqref{Poisson} and momentum conservation.
Inserting Eq.~\eqref{Triggerrelation} into \eqref{growthequation}, one obtains a relation between $\mu_{\infty}(a)$ and $\gamma$,
\begin{equation}
\mu_{\infty}=\frac{2}{3}\Omega_{m}^{\gamma-1}\left[\Omega_{m}^{\gamma}+2+\frac{H^{\prime}}{H}+\gamma\frac{\Omega_{m}^{\prime}}{\Omega_{m}}+\gamma^{\prime} \textnormal{ln} \, (\Omega_{m})   \right] \,,
\label{muintermsofgrowth}
\end{equation}
where we allowed $\gamma$ to be time dependent for generality.
Given a particular choice of $\gamma$, the functional form of $\mu_{\infty}$ can then be obtained from Eq.~\eqref{muintermsofgrowth}.
However, as $\gamma$ can only be used to determine $\mu_{\infty}$, one must separately parameterize the gravitational slip $\eta_{\infty}$ (and some additional specifications are required for a relativistic completion~\cite{Lombriser:2011tj,Lombriser:2013aj,Lombriser:2015cla}).
One can then reconstruct a covariant theory that gives rise to the particular choices of $\gamma$ and $\eta_{\infty}$.
This allows to directly examine what kind of theories
can be associated with an observational departure from GR in $\gamma$.

In our example, we set for simplicity $\eta_{\infty}=1$ as in GR.
This implies that, with $\alpha_{T}=0$,  $\alpha_{M}=0$ or $\alpha_{B}=\alpha_{M}$. We choose the second condition. With this choice we have that $M^{2}=1/\mu_{\infty}$ and we fix $\alpha_{K}$ such that $c_{s}^{2}=1$.
We shall reconstruct a theory which gives rise to a constant deviation in the growth index from the GR value of $\gamma \approx 0.55$.  
The value for $\gamma$ needs to be chosen such that the stability condition $\alpha >0$ is satisfied and so we choose $\gamma=0.4$ for this purpose. In fact, 
the theoretical stability of the theory requires $0.35 \lesssim \gamma \lesssim 0.55 $, preferring enhanced growth of structure, with any value chosen outside this range leading to $\alpha<0$. 
As long as the theoretical conditions are satisfied then it is straightforward to apply the reconstruction and obtain a covariant theory for any numerical value for $\gamma$.

The corresponding model is illustrated in Fig.~\ref{Growthrateandmuandgamma}. As we have chosen a rather large departure from $\Lambda$CDM the reconstructed theory displays a somewhat unnatural behavior with a potential that is negative and substantial contributions from the kinetic and Galileon terms in order to maintain the background expansion history. Therefore, even with this seemingly simple parameter it is quite possible that exotic regions of the space of theories are being explored when it deviates from its concordance value.

%%%%%%%%%%%%%%   Weak Gravity    %%%%%%%%%%%%%%%%

\subsection{Weak gravity}
\label{sec:IIIF}

%%% FIGURE WEAK GRAVITY 1 %%%
\begin{figure*}
\resizebox{0.495\textwidth}{!}{
\includegraphics{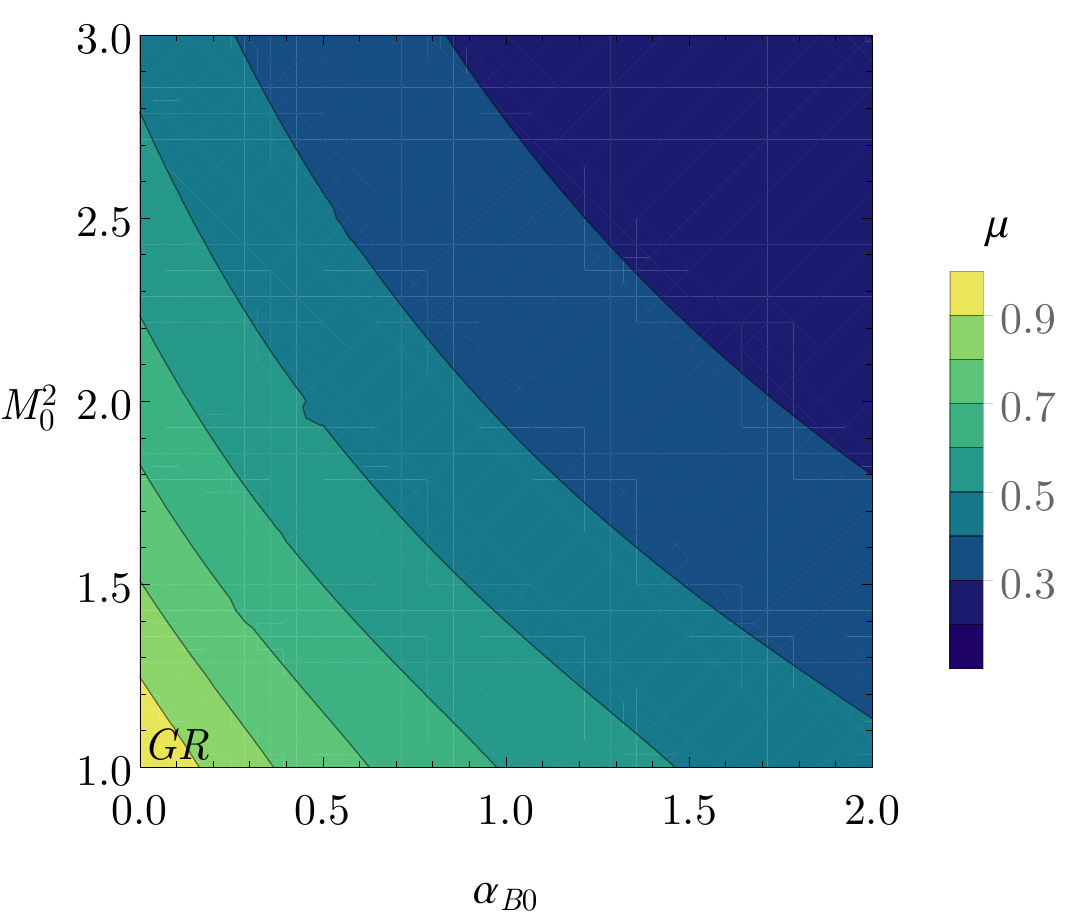}
}
\resizebox{0.44\textwidth}{!}{
\includegraphics{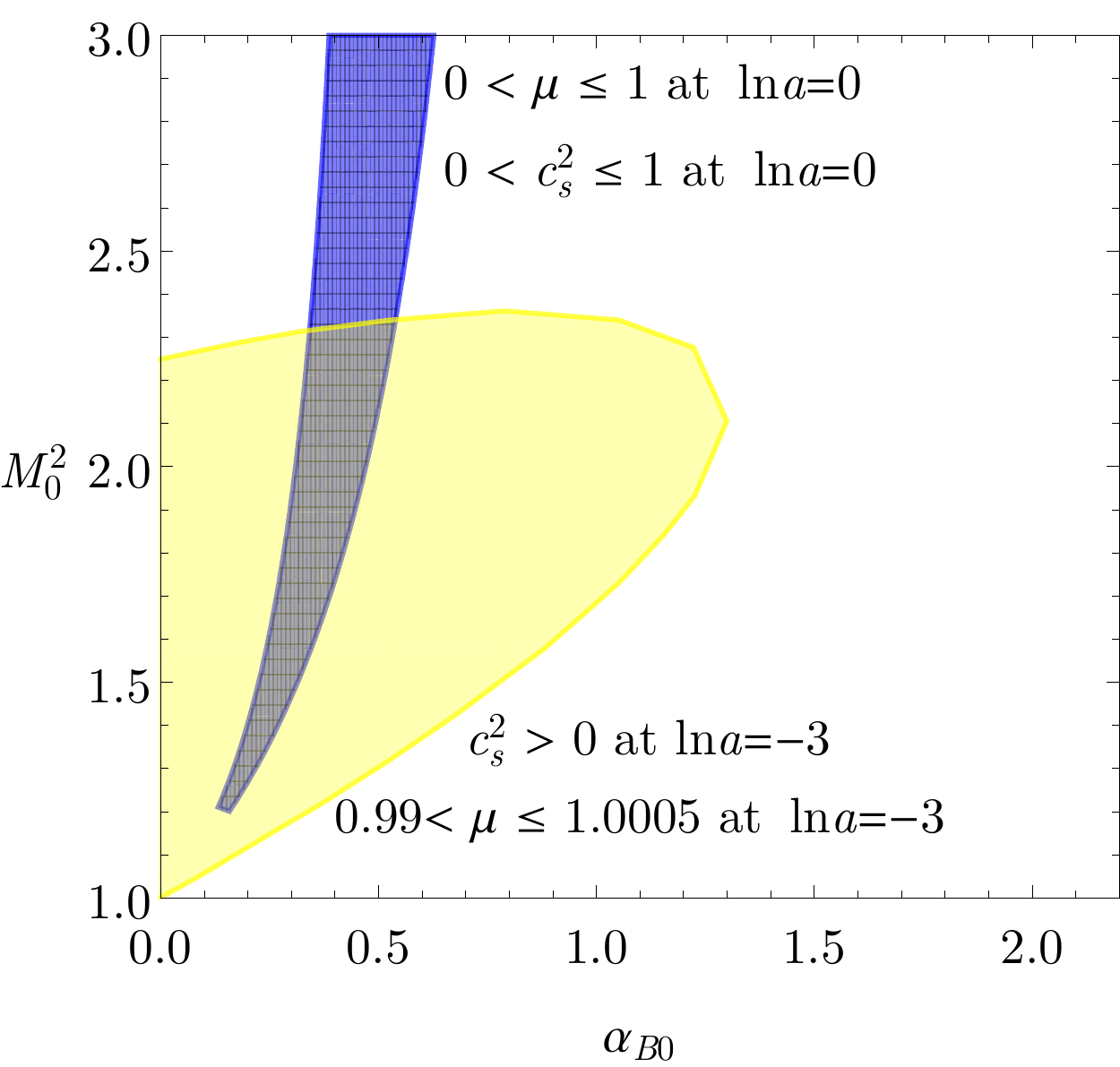}
}
\caption{\emph{Left}: Contour plot in the space of $M_{0}^{2}$ and $\alpha_{B0}$ displaying the regions that allow for a weakened growth of structure with $0<\mu_0<1$ today. \emph{Right}: The dark strip indicates the region of EFT parameter space that allows for a weakening of growth with a positive, sub-luminal soundspeed at redshift zero. After imposing the past boundary conditions $\mu = 1$ and $c_{s}^{2}>0$ at $\ln a=-3$ indicated by the lighter yellow region it is possible to reconstruct a viable covariant model from any point in the intersecting region. We have ensured that the chosen point used for the reconstruction in Fig.~\ref{Weakgravityplots} satisfies $c_{s}^{2}>0$ for all time.
}
\label{ContourplotsWG}
\end{figure*}
%%% FIGURE WEAK GRAVITY 1 %%%

%%% FIGURE WEAK GRAVITY 2 %%%
\begin{figure*}
\resizebox{0.499\textwidth}{!}{
\includegraphics{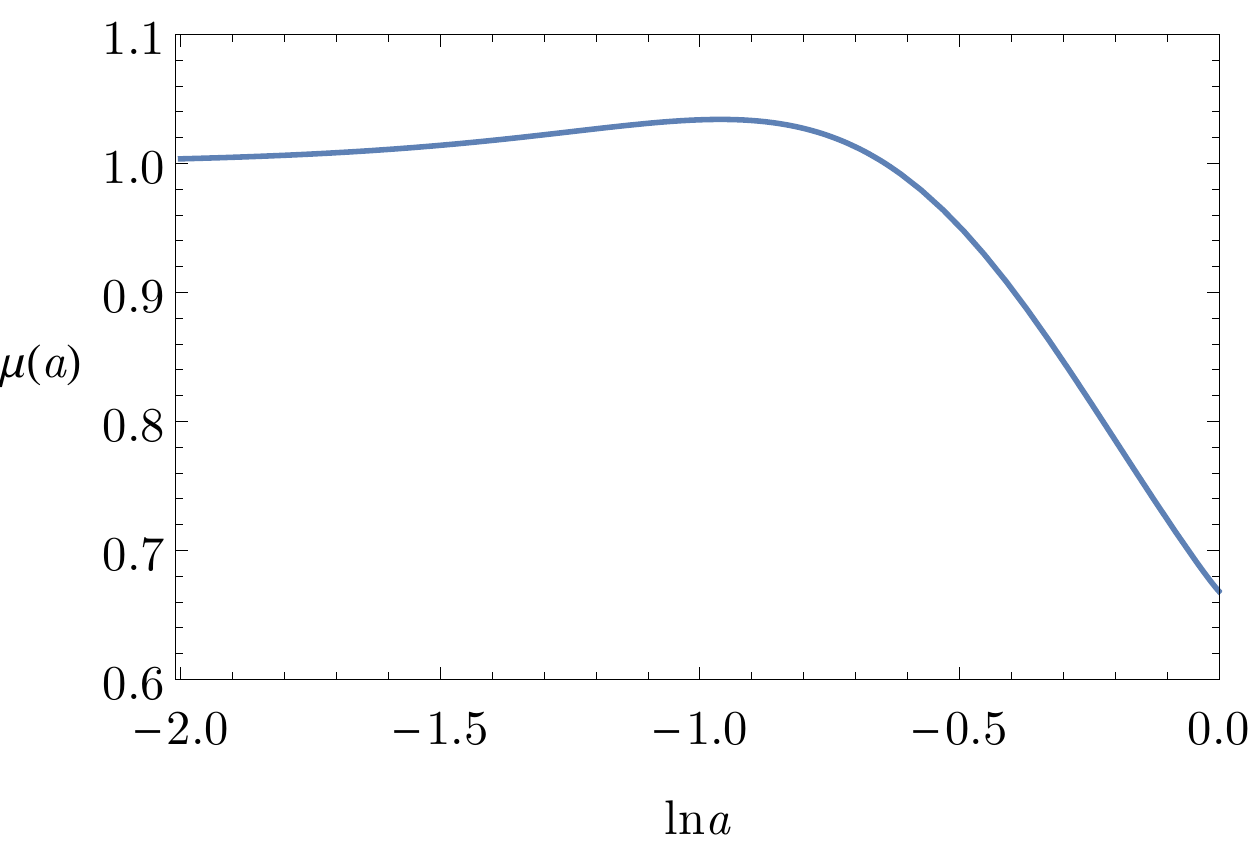}
}
\resizebox{0.46\textwidth}{!}{
\includegraphics{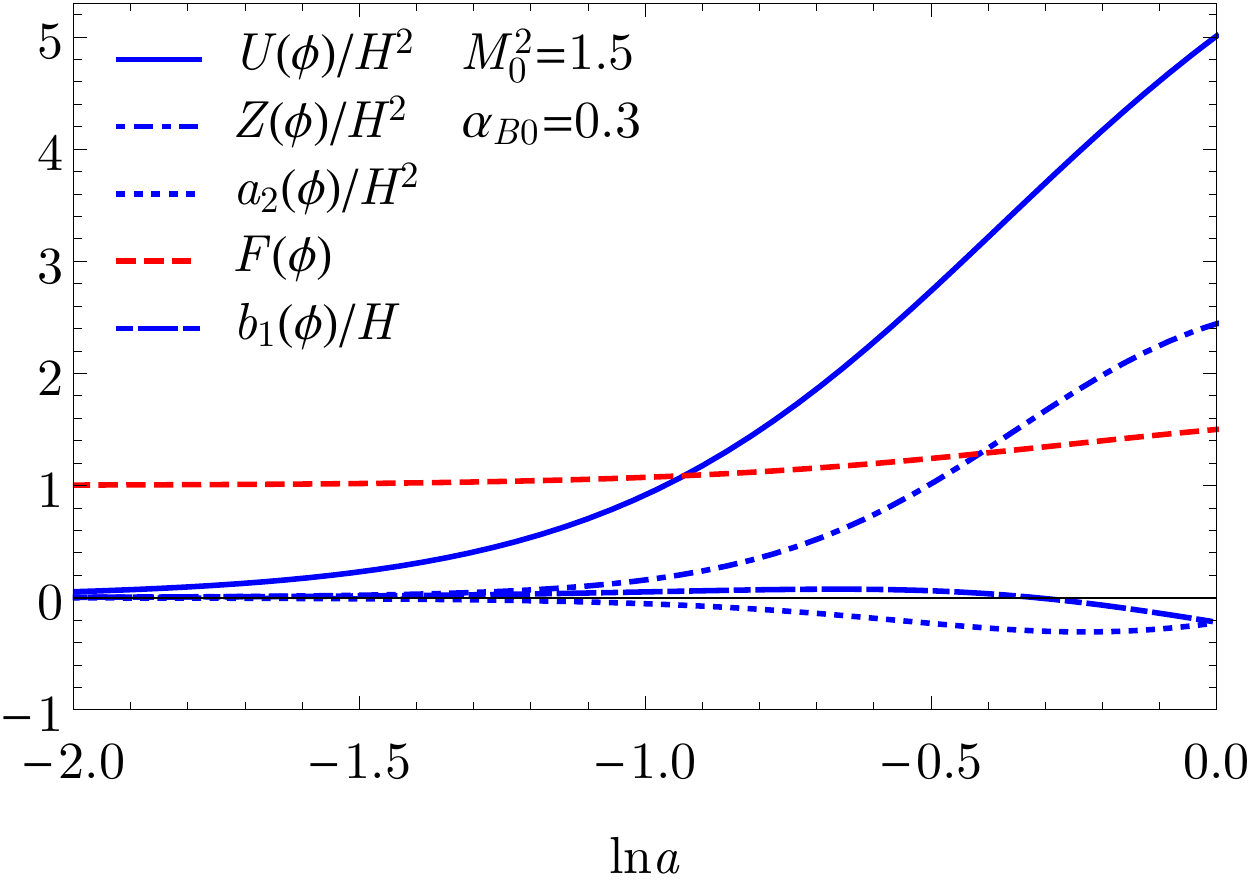}
}
\caption{\emph{Left}: The behavior of the deviation from the Poisson's equation over time for the model in Sec.~\ref{sec:IIIF}, where one may identify a dynamical $G_{\rm eff}\equiv\mu$.
There is a characteristic period of enhanced growth at $\textnormal{ln}\, a\approx -0.96$ before entering an epoch of weakening of the growth persisting today.  \emph{Right}: A reconstructed scalar-tensor theory that exhibits a weakening of the growth of structure (``weak gravity'') with $\alpha_{T}=0$, which satisfies the stability requirements and past boundary conditions. It is essentially a Brans-Dicke type model with a potential and standard kinetic term along with small contributions from the k-essence and cubic terms.}
\label{Weakgravityplots}
\end{figure*}
%%% FIGURE WEAK GRAVITY 2 %%%

Typically scalar-tensor theories exhibit an enhanced growth of the matter density fluctuations relative to $\Lambda$CDM, with Brans-Dicke gravity being a simple example~\cite{PhysRev.124.925}. More precisely, they lead to a modification such that $\mu>1$ in Eq.~\eqref{Poisson}. However, it is possible that modifications arise such that one obtains a weaker growth of structure, or \emph{weaker gravity}, with $\mu<1$.
This scenario has recently received some attention~\cite{Piazza:2013pua, Tsujikawa:2015mga, Perenon:2015sla, DAmico:2016ntq, Linder:2018jil},
particularly in the context of potential tensions in the cosmological data~\cite{Abbott:2017wau, Amon:2017lia}.

In this section we demonstrate how one may use the reconstruction to derive a stable scalar-tensor theory of weak gravity for a particular parameterization of the EFT functions with $\alpha_T=0$.

We begin by choosing the parameterization of the Planck mass $M^{2}$ as
\begin{equation}
  M^{2}=1+(M_{0}^{2}-1)\frac{\Omega_{\Lambda}(a)}{\Omega_{\Lambda 0}} \, ,
  \label{PlanckmassParam}
\end{equation}
where $M_{0}^{2}$ is the value of the Planck mass today.
The particular choice of Planck mass evolution when $M_{0}^{2}>1$ is a priori suggestive of weak gravity as $M^2$ appears in the denominator of Eq.~\eqref{muinf} such that the increasing Planck mass with time leads to a decreasing $\mu$ if fixing the other EFT parameters.
However, there is still a great deal of freedom in choosing numerical values for $M_{0}^{2}$ and the evolution of the remaining $\alpha_i$.
For instance, it may be the case that the evolution in $\alpha_{B}$ is enough to compensate for the weakened growth effect and give rise to an enhancement instead.
For our example, we adopt the functional form of $\mathcal{B}$ in Sec.~\ref{sec:IIIA} with $q_{i}=q=1$ for the parameterization of the $\alpha_B$ function
and we set $\alpha_{K}=0$ for simplicity and
to easily guarantee that the stability condition $\alpha>0$ is satisfied.
As previously mentioned, $\alpha_{K}$ only becomes relevant on scales comparable to the horizon and so the requirement that $\mu<1$ is independent of the choice of $\alpha_{K}$.
Parameter values for $M_0^2$ and $\alpha_{B0}$ are then chosen to ensure that the condition $c_{s}^{2}>0$ is satisfied.

We explore the viable regions of parameter space producing a given $\mu_0\equiv\mu(z=0)$ in the left-hand panel of Fig.~\ref{ContourplotsWG}.
One can easily identify
a large region
that allows for weak gravity with $0<\mu_0<1$ when $M_{0}^{2}>1$
while remaining stable and having the Planck mass return to its bare value in the past by construction.
All of these requirements severely restrict the allowed model space.
In fact, we find that within the particular parameterization adopted here, a period of enhanced growth in the past is required in order for all of these criteria to be satisfied.

We explore this circumstance in more detail in the right-hand panel of Fig.~\ref{ContourplotsWG}.
For this purpose, we allow for a small period of enhanced growth in the past
at
$\mathcal{O}(10^{-4})$, which allows one to find an overlap of stable parameter choices that also yield weak gravity at late times.
Increasing this value causes the viable parameter regions to overlap at an even greater extent.
Restricting parameters to an upper bound of exactly unity instead eliminates any overlap.

A suitable parameter choice that satisfies all of the requirements described here is $M_{0}^{2}=3/2$ and $\alpha_{B0}=0.3$ and we checked that for this choice the soundspeed remains positive at all times in the past.
The left-hand panel of Fig.~\ref{Weakgravityplots} displays the evolution of the gravitational coupling through time with this choice of EFT parameters. One can clearly identify a period of enhanced growth which peaks around $\textnormal{ln} \, a\approx -0.96$ with $\mu\approx 1.03$ before decaying
and producing weak gravity with $\mu \approx 0.65$ at redshift $z=0$.

Once given the choice of EFT parameters it is straightforward to implement them in the reconstruction and obtain a stable scalar-tensor theory that exhibits a weakening of growth of structure with $\alpha_{T}=0$.
The corresponding model is illustrated in the right-hand panel of  Fig.~\ref{Weakgravityplots}. 
The evolution of $U(\phi)$ mimics that of a cosmological constant, but as $\Lambda \sim M^{2}H^{2}$ it is enhanced relative to its $\Lambda$CDM behavior due to the increase of the Planck mass over time.
This is similar to the behavior observed in Sec.~\ref{sec:IIID}.
The Planck mass also determines the evolution of $F(\phi)$ which increases over time.
The behavior of $b_{1}(\phi)$ is determined by the combination
$\alpha_{M}-2\alpha_{B}$. The braiding term is sub-dominant at early times, but becomes important at late times,
where it contributes to drive $b_{1}(\phi)$ negative. 
There is also a small negative k-essence term $a_{2}(\phi)$ that is comparable in magnitude to $b_{1}(\phi)$. 

Bear in mind that different choices of $q_{0}$, a non-zero $\alpha_{K}$ or a parameterization in terms of $\alpha_{M}$ rather than $M^{2}$ impacts the form of the theory.
However,
it is primarily sensitive to significant changes in the amplitudes of parameters as discussed in the Appendix, and one does not have much freedom in increasing the amplitude of $\alpha_{B}$ while keeping the theory stable (Fig.~\ref{ContourplotsWG}).
Finally, note that our weak gravity model differs from Ref.~\cite{Linder:2018jil} as
$\alpha_{M}\neq\alpha_{B}$,
thus exhibiting a non-vanishing gravitational slip.
More work is necessary to understand what general conditions need to hold in order to obtain a stable scalar-tensor theory the exhibits a weakened growth of structure and $\alpha_{T}=0$.

\subsection{Reconstruction from inherently stable parameterizations}
\label{sec:IIIG}

%%% FIGURE STABLE PARAM %%%
\begin{figure}[t]
\resizebox{0.485\textwidth}{!}{
\includegraphics{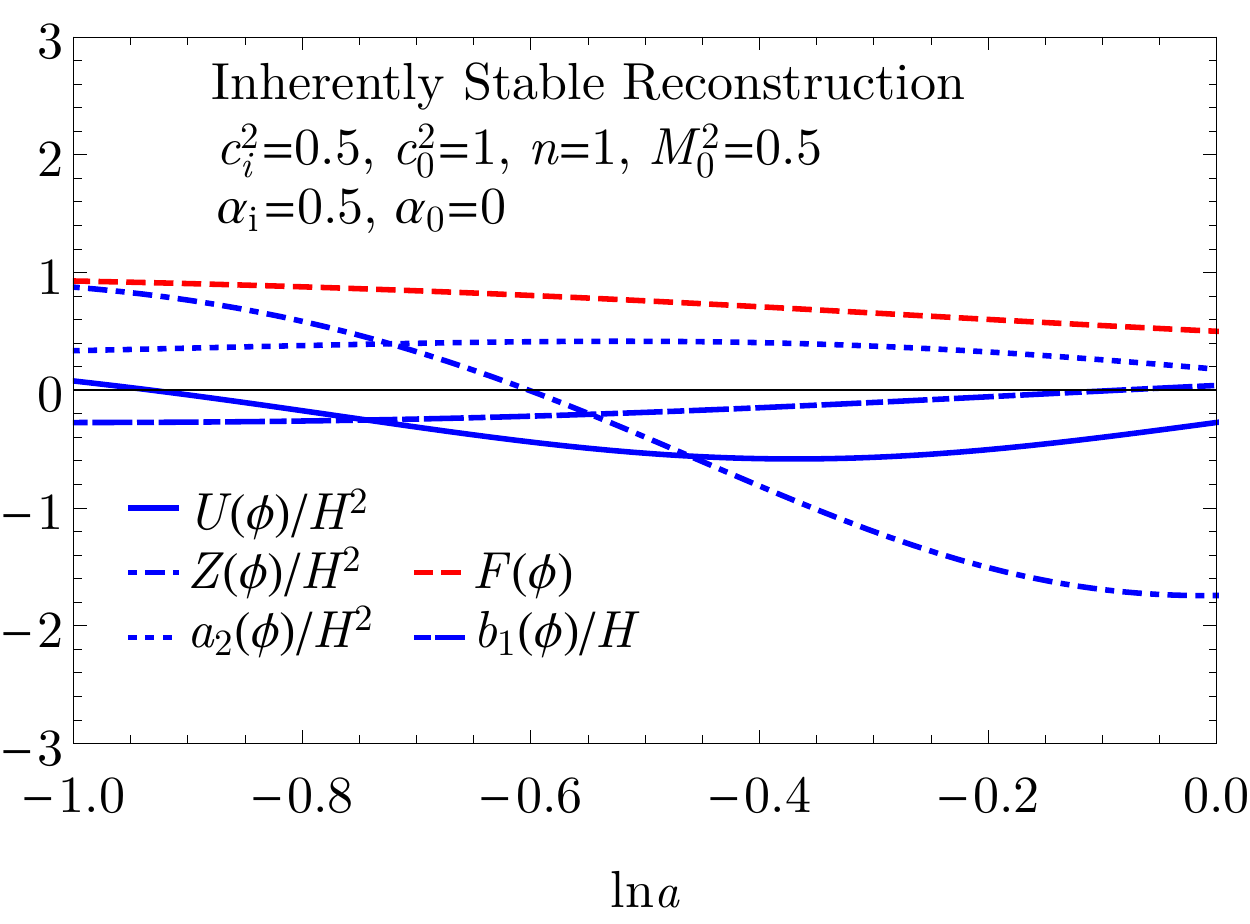}
}
\caption{Reconstructed scalar-tensor theory from a direct parametrization of the stability functions $c_{s}^{2}>0$, $\alpha>0$ and $M^{2}>0$ with $\alpha_T=0$.}
\label{ISRec}
\end{figure}
%%% FIGURE STABLE PARAM %%%

Throughout this work it has been necessary to check that the reconstructed theories obey the stability constraints in Eqs.~\eqref{Stabcriteria1} and \eqref{Stabcriteria2}.
This is due to the function space spanned by the basis of $\alpha_i$, or equivalently the coefficients in the EFT action in Eqs~\eqref{eq:s01} and \eqref{eq:s2}, not being a priori stable.
As discussed in Sec.~\ref{sec:stableparam}, rather than cumbersomely checking that these stability criteria are satisfied for a particular parameterization,
one may instead consider discarding the $\alpha_i$ functions in favor of another parameterization that automatically satisfies the stability requirements.
Therefore, any observational constraints
will by definition be restricted to
a theory space that obeys the no-ghost and no-gradient instability conditions.
We introduced such an inherently stable basis in Sec.~\ref{sec:stableparam}.

We shall now briefly present a reconstruction from this basis. 
For this purpose we adopt the functional forms
\begin{eqnarray}
c_{s}^{2} & = &c_{i}^{2}+(c_{0}^{2}-c_{i}^{2})a^{n}  \,,  \\  \alpha & = &\alpha_{i}+(\alpha_{0}-\alpha_{i})a^{n} \,,
\end{eqnarray}
where the constants $c_{i}^{2}$ and $\alpha_{i}$ are initial conditions for the soundspeed and the kinetic term respectively (defined for the limit $a\rightarrow0$) 
whereas $c_{0}^{2}$ and $\alpha_{0}$ set their values today.
Each value should be chosen such that $\alpha,c_{s}^{2}>0$ $\forall a$.
For the Planck mass we adopt the parameterization in Eq.~\eqref{PlanckmassParam}.

In Fig.~\ref{ISRec} we illustrate a reconstructed theory with a $\Lambda$CDM background, an increasing soundspeed as well as decaying kinetic term and Planck mass.
More specifically, we set $c_{i}^{2}=0.5$, $c_{0}^{2}=1$, $\alpha_{i}=0.5$, $\alpha_{0}=0$, $M_{0}^{2}=0.5$, and $n=1$.
Although the reconstructed terms seem somewhat exotic, for example the potential is very different to its $\Lambda$CDM behavior despite the concordance background evolution, by construction the model is guaranteed to be stable.

%%%%%% CONCLUSIONS %%%%%%

\section{Conclusions}
\label{sec:conclusions}

%Obtaining
Finding a natural explanation for the observed late-time accelerated expansion of our Universe
continues to be
a significant
challenge in cosmology.
It is
therefore important
that efficient methods are devised with the aim of connecting cosmological observables with the wealth of proposed theories to obtain a deeper understanding of the underlying physical mechanism driving the expansion.
These efforts may furthermore give crucial insights into the persistent issues related to the reconciliation of quantum field theory with general relativity.  

The effective field theory of dark energy provides a useful tool for studying the dynamics of cosmological perturbations of a large family of scalar-tensor theories in a unified framework. 
Many of the upcoming surveys of the large-scale structure plan to utilize this formalism to constrain the freedom in modified gravity and dark energy phenomenology~\cite{Laureijs:2011gra, Amendola:2012ys, Ivezic:2008fe}.
It is therefore 
crucial to be able to connect
any observational
constraints to the underlying space of scalar-tensor theories, which in turn can be connected to more fundamental theories of gravity. 

Recently we have developed a reconstruction method that maps from a set of EFT functions to the family of Horndeski theories degenerate at the level of the background and linear perturbations \cite{Kennedy:2017sof}.
In this paper we apply this mapping to a number of examples. 
These include the comparison of the resulting action when one utilizes two frequently adopted phenomenological parameterizations for the EFT functions to study the effects of dark energy and modified gravity at late times.
We find that changing between the two parameterizations has a small effect on the general form of the underlying theory, although certain terms can be enhanced relative to others. The underlying theory is instead more sensitive to the amplitudes of the different EFT functions.

Of particular interest is the reconstruction of a model that exhibits minimal self-acceleration. The reconstructed scalar-tensor theory
possesses the minimum requirements on the evolution of the Planck mass for self-acceleration from a modification of gravity
consistent with a propagation speed of gravitational waves equal to that of light.
It is a useful model to test for the next generation of surveys, as it acts as a null-test for self-acceleration from modified gravity. 

We also examine models that exhibit a linear shielding mechanism to hide the gravitational modifications in the large-scale structure.
Although the simplest models require a non-vanishing $\alpha_{T}$, it is worth bearing in mind that the stringent constraint on the speed of gravity with $\alpha_T=0$ only applies
at low redshifts and may also involve scale dependence~\cite{Battye:2018ssx} for more general theories.
While the constraints in the space of the EFT functions for linear shielding to operate seem rather complicated, using the reconstruction we find there are generic Horndeski theories that exhibit this effect.

We furthermore
provide a direct connection between
various parameterizations that exist in the literature and the corresponding underlying theories.
For example, we reconstruct theories from a phenomenological parameterization of the modified Poisson equation and gravitational slip as well as from the growth-index parameter. One can use these reconstructions to connect constraints arising from such parameterizations with viable Horndeski models. 
We also apply the reconstruction to obtain a theory that exhibits a weakening of the present growth of structure relative to $\Lambda$CDM, i.e., a weak gravity model, a possibility that may ease potential tensions in the growth rate at low redshift~\cite{Abbott:2017wau, Amon:2017lia}.

Finally, we propose an alternative parameterization basis for studying dark energy and modified gravity models which is manifestly stable. These are the Planck mass, the dark energy soundspeed, the kinetic energy of the scalar field and a braiding amplitude as the new basis of EFT functions. 
Any constraints placed on these physical parameters are guaranteed to correspond to healthy theories. It is no longer necessary to
perform
separate and cumbersome stability checks on sampled theories when using this basis.

Many further applications of the reconstruction remain to be addressed, the development of which will be the subject of future work.

%%%%%%%%% APPENDIX %%%%%%%%%%%%%

\appendix
\section{Effect of varying the parameterization on the underlying theory}

%%% FIGURE APP 1 %%%
\begin{figure*}
\resizebox{0.495\textwidth}{!}{
\includegraphics{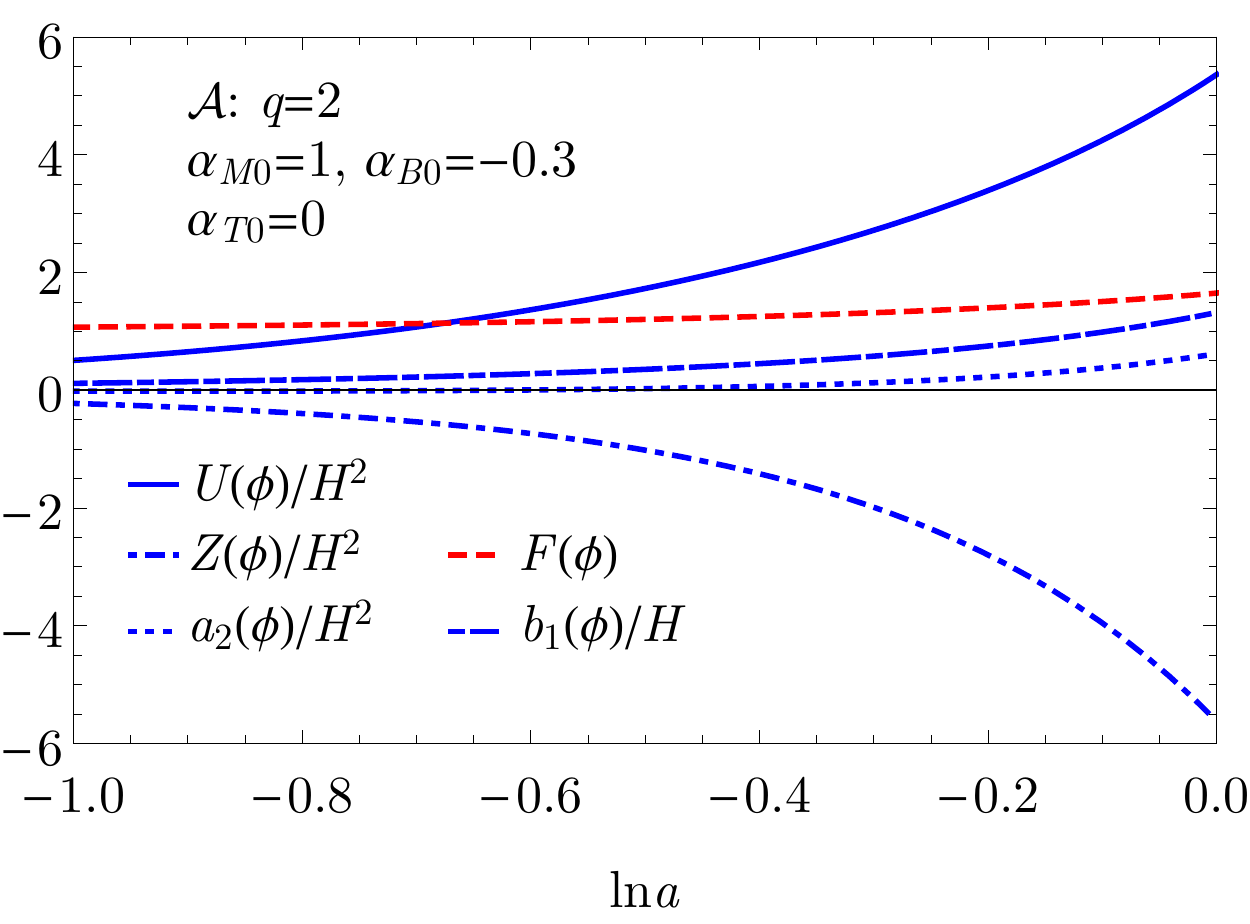}
}
\resizebox{0.495\textwidth}{!}{
\includegraphics{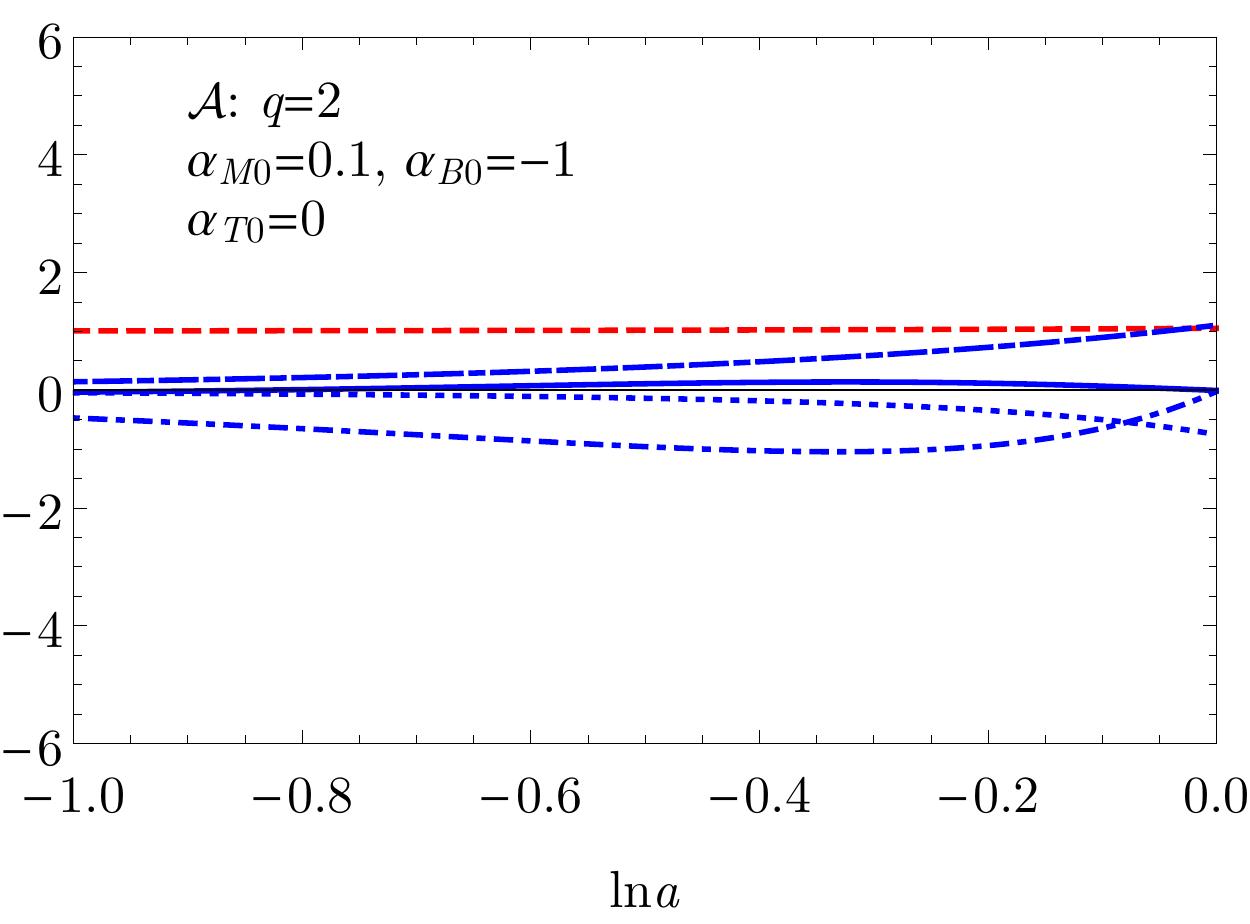}
}
\label{ParAq2EnhancedM} \label{ParAq2EnhancedB}
\caption{The effect of varying parameter values in a parameterization of EFT functions on the reconstructed scalar-tensor theory for a model with a dominant Planck mass evolution $\alpha_{M}$ (\emph{left panel}) and a model with a dominant braiding term $\alpha_{B}$ (\emph{right panel}).
Note that $\alpha_{M0}$ and $\alpha_{B0}$ are of opposite sign to satisfy the stability requirements.
In the right-hand panel where $\alpha_{B}$ dominates, the cubic Galileon term $b_1$ is the most prevalent modification as the potential and quintessence terms decay to zero. There is also a non-negligible contribution from the k-essence term. On the contrary, a dominating $\alpha_{M}$ leads to a large potential and quintessence kinetic term, with smaller contributions from the cubic and k-essence terms. 
}
\label{Enhancedplots}
\end{figure*}
%%% FIGURE APP 1 %%%

%%% FIGURE APP 2 %%%
\begin{figure*}
\resizebox{0.495\textwidth}{!}{
\includegraphics{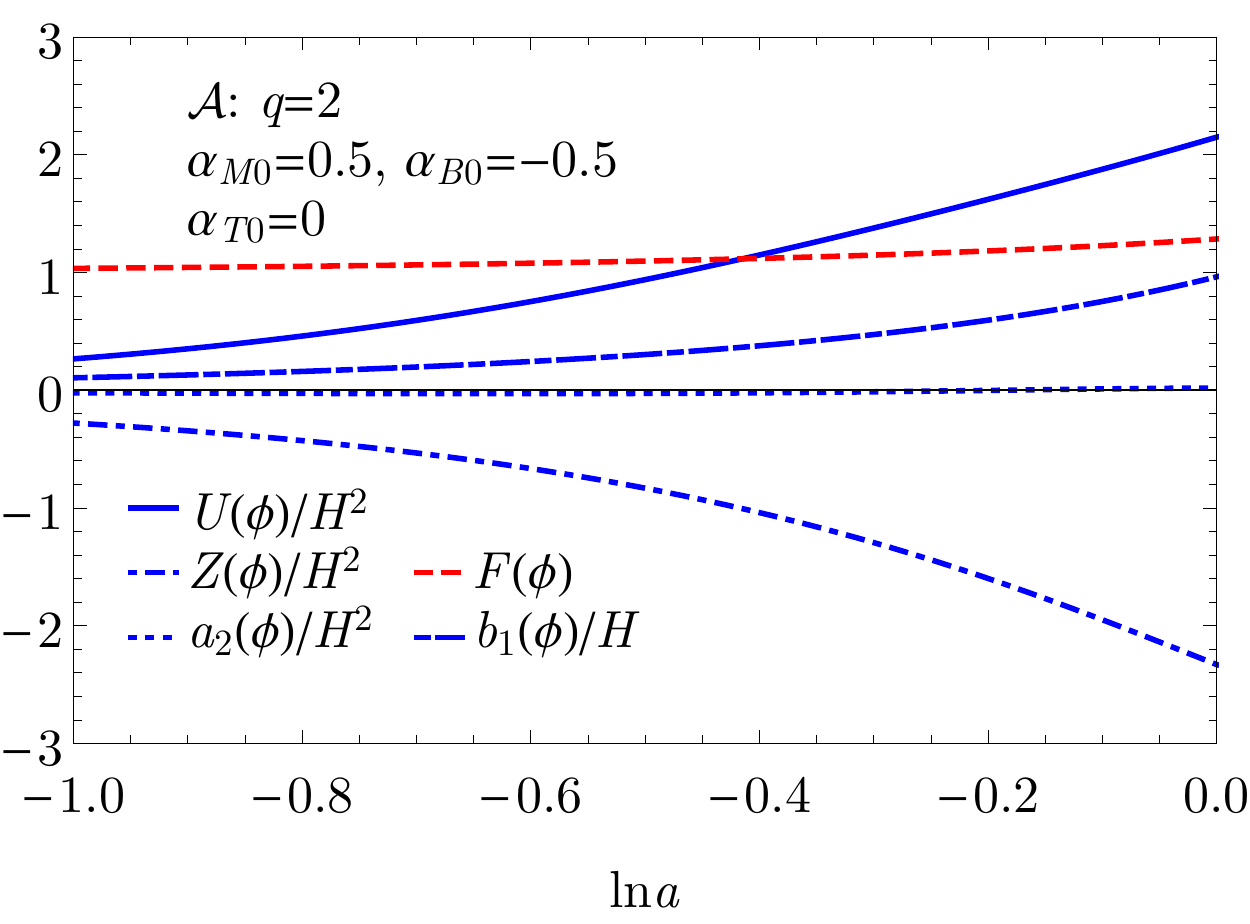}
}
\resizebox{0.495\textwidth}{!}{
\includegraphics{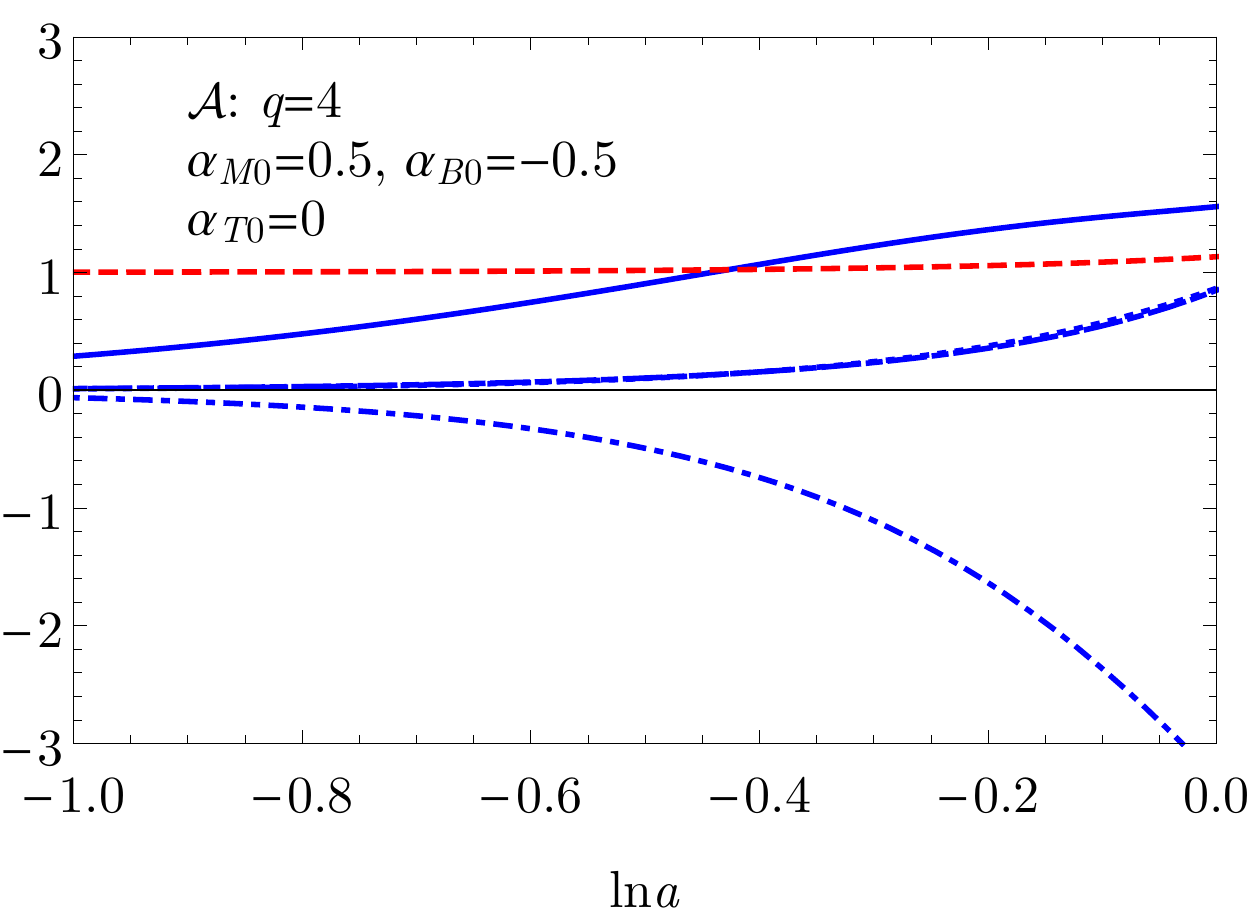}
}
\resizebox{0.495\textwidth}{!}{
\includegraphics{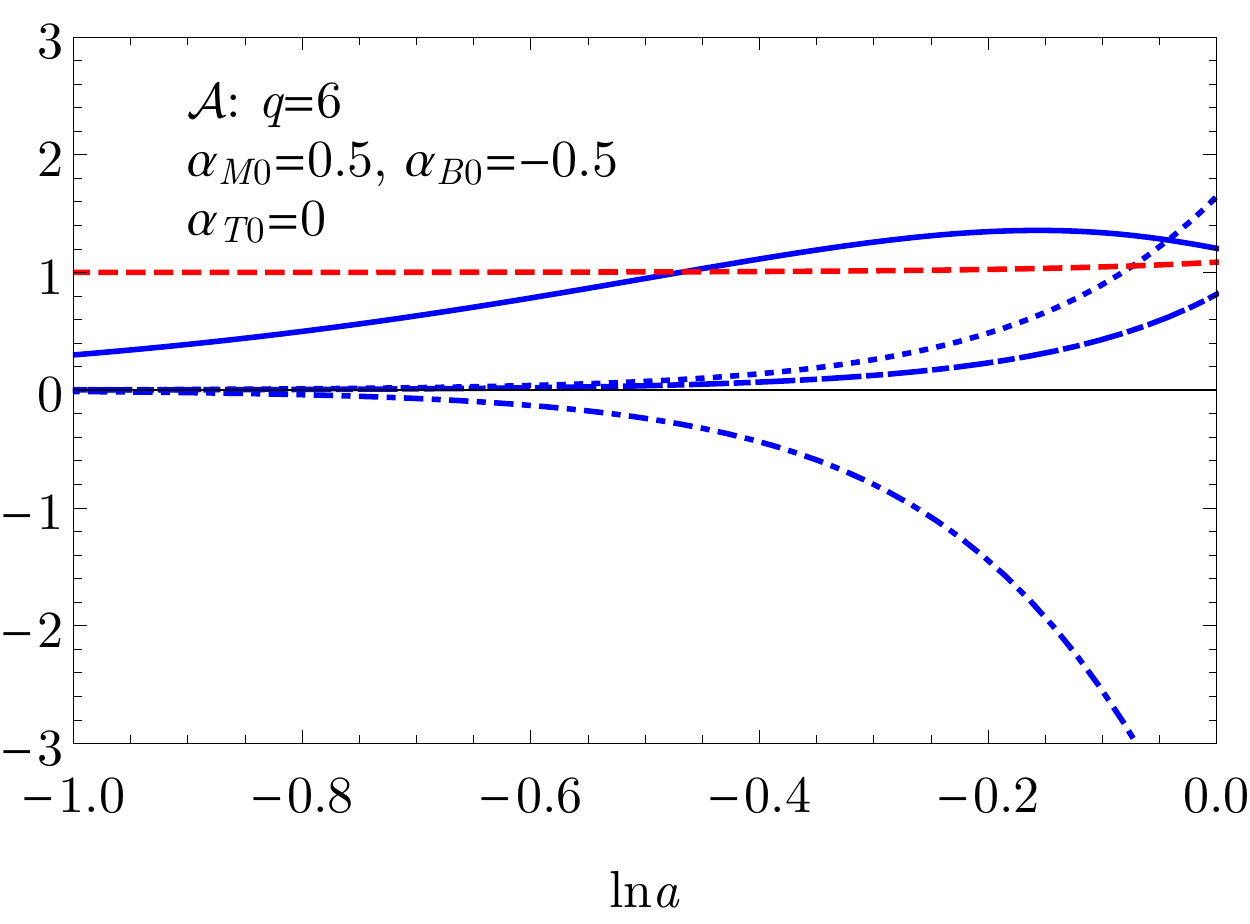}
}
\resizebox{0.495\textwidth}{!}{
\includegraphics{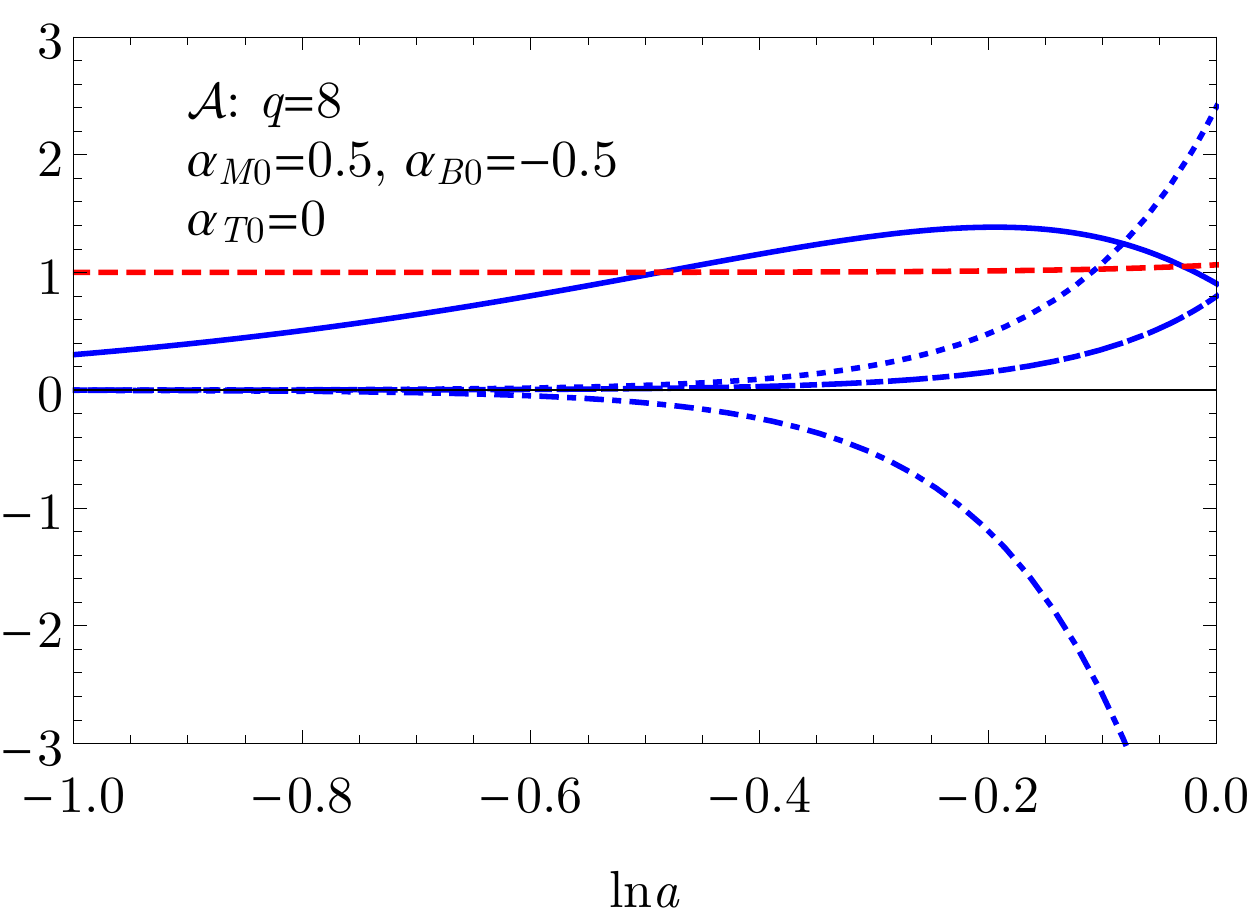}
}
\label{ParAq2}\label{ParAq4}\label{ParAq6}\label{ParAq8}
\caption{The effect of varying the powers $q$ in the parameterization on the underlying theory.
It is apparent that with this choice of $\alpha_i$ functions
every term in the reconstruction becomes relevant.
Modifications are
suppressed at high redshift with increasing power, with a steepening
at low redshifts.
For this choice of amplitudes, the k-essence term
is particularly sensitive, increasing from zero to dominate over the potential for large $q$. The standard kinetic term and potential become more negative at $z=0$ for larger powers.
This is in contrast to the cubic term $b_{1}(\phi)$, which
remains relatively unaffected by this
alteration in
the parameterization.
}
\label{VaryingSlope}
\end{figure*}
%%% FIGURE APP 2 %%%

%%% FIGURE APP 3 %%%
\begin{figure*}
\resizebox{0.485\textwidth}{!}{
\includegraphics{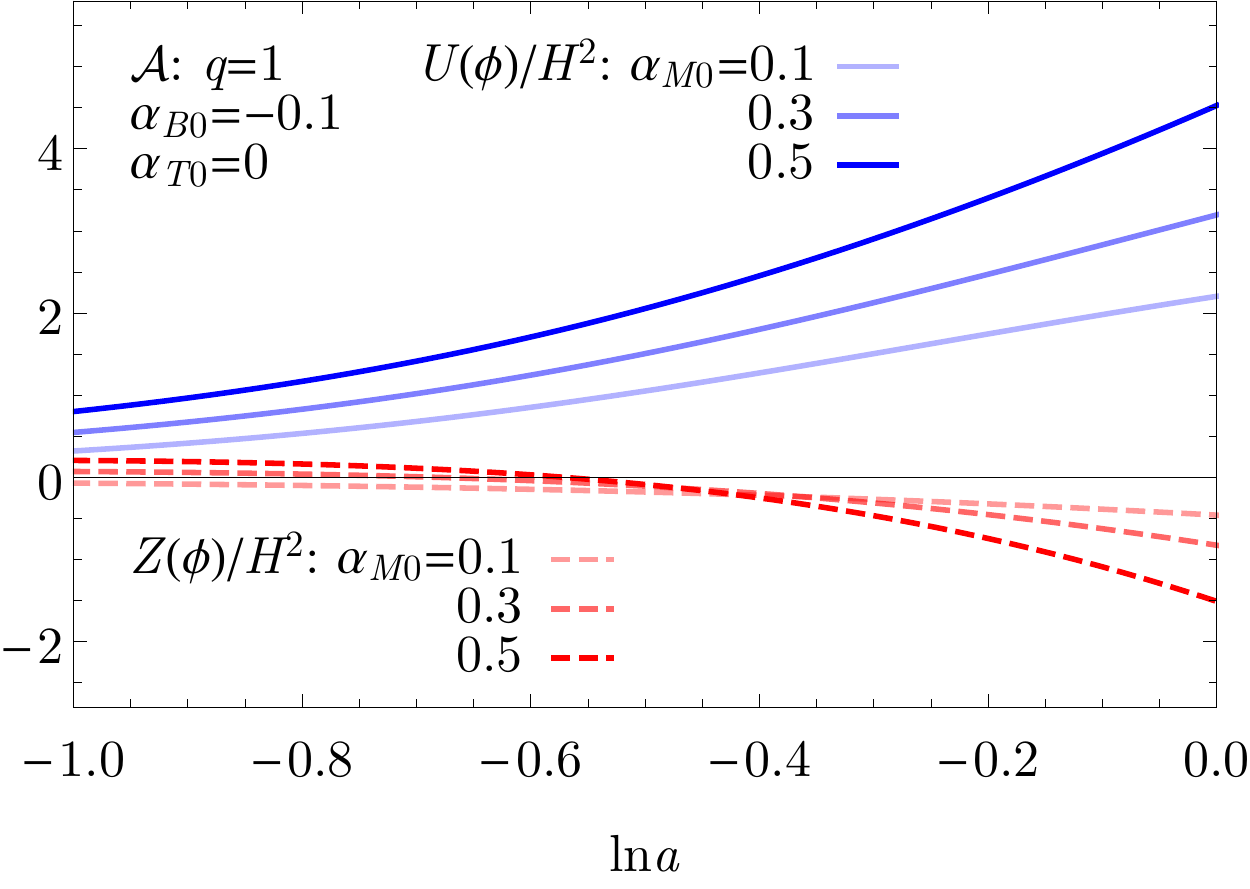}
}
\resizebox{0.499\textwidth}{!}{
\includegraphics{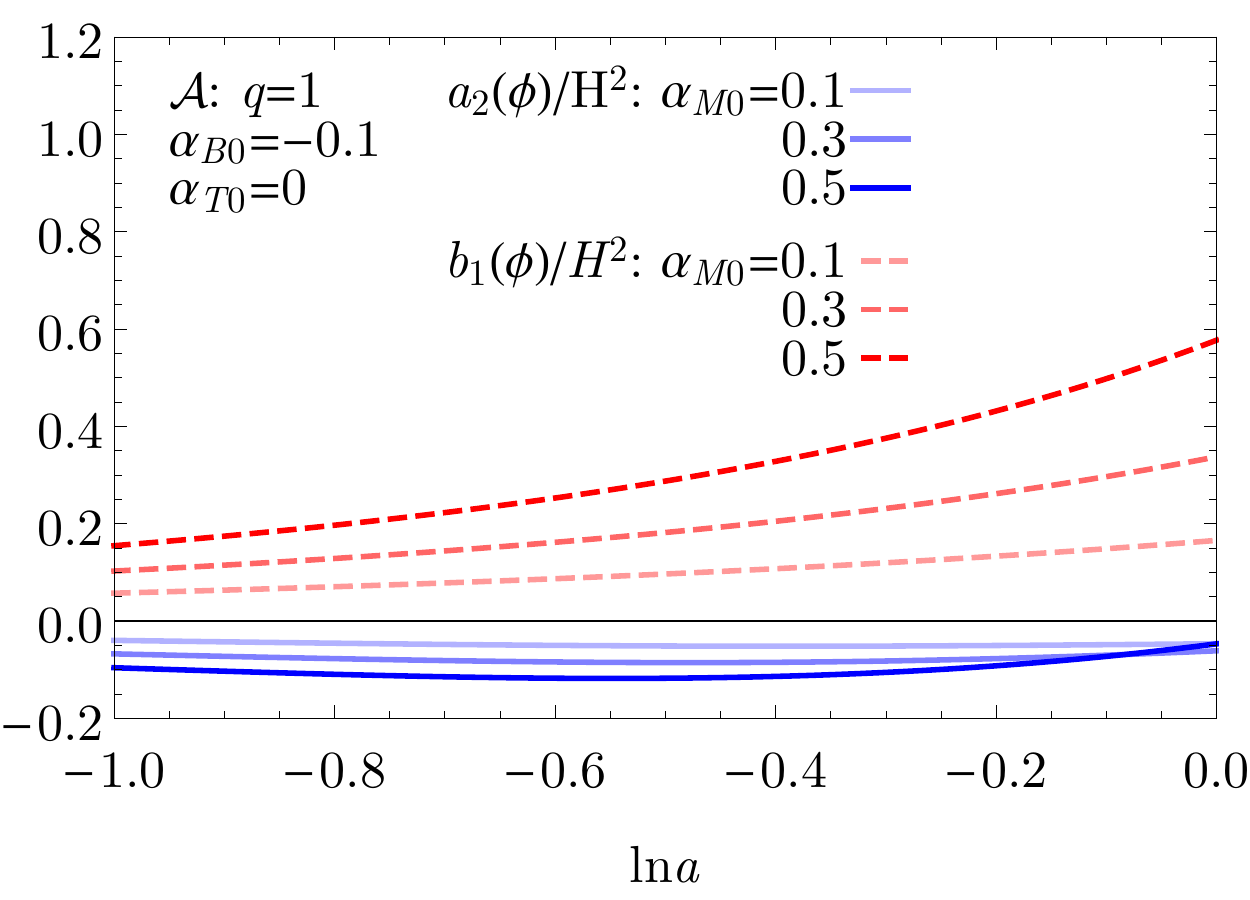}
}
\resizebox{0.485\textwidth}{!}{
\includegraphics{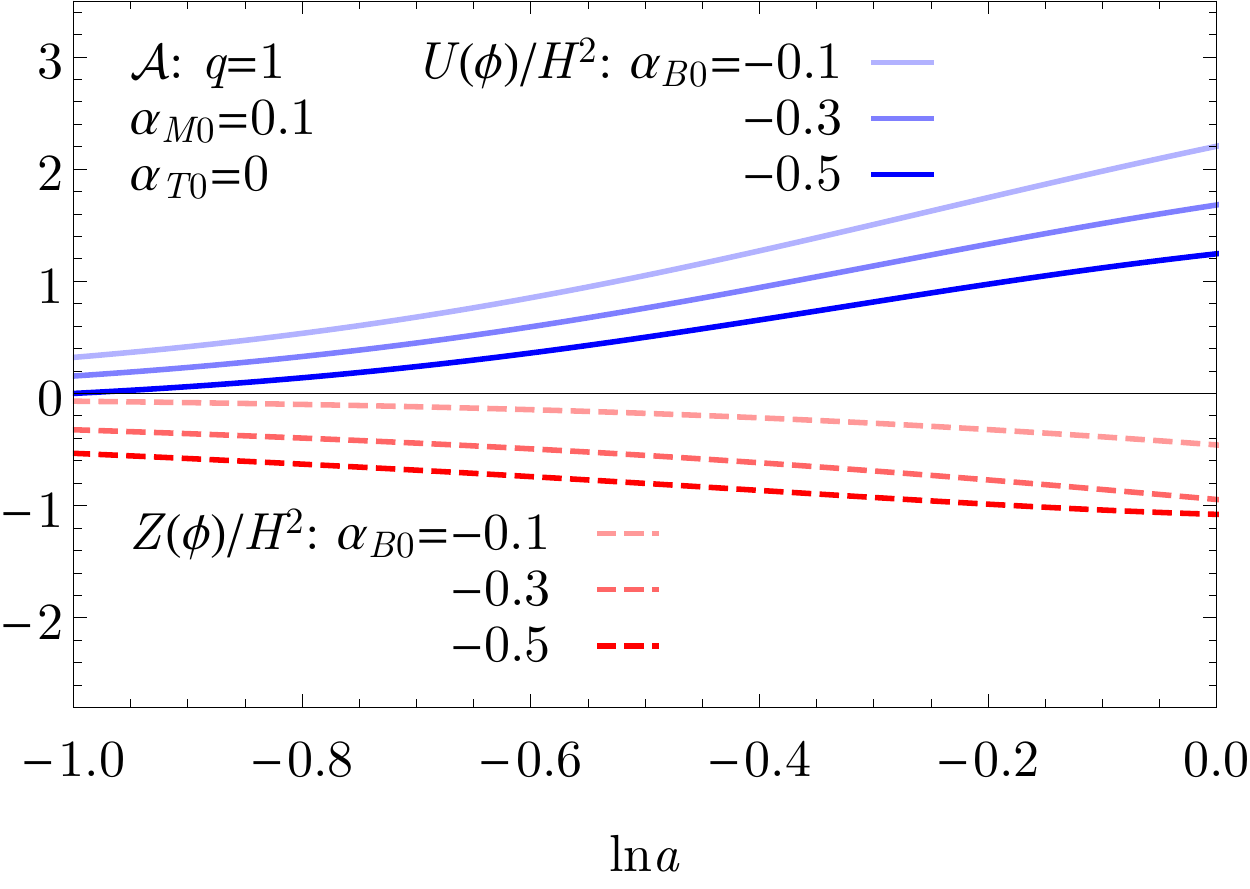}
}
\resizebox{0.499\textwidth}{!}{
\includegraphics{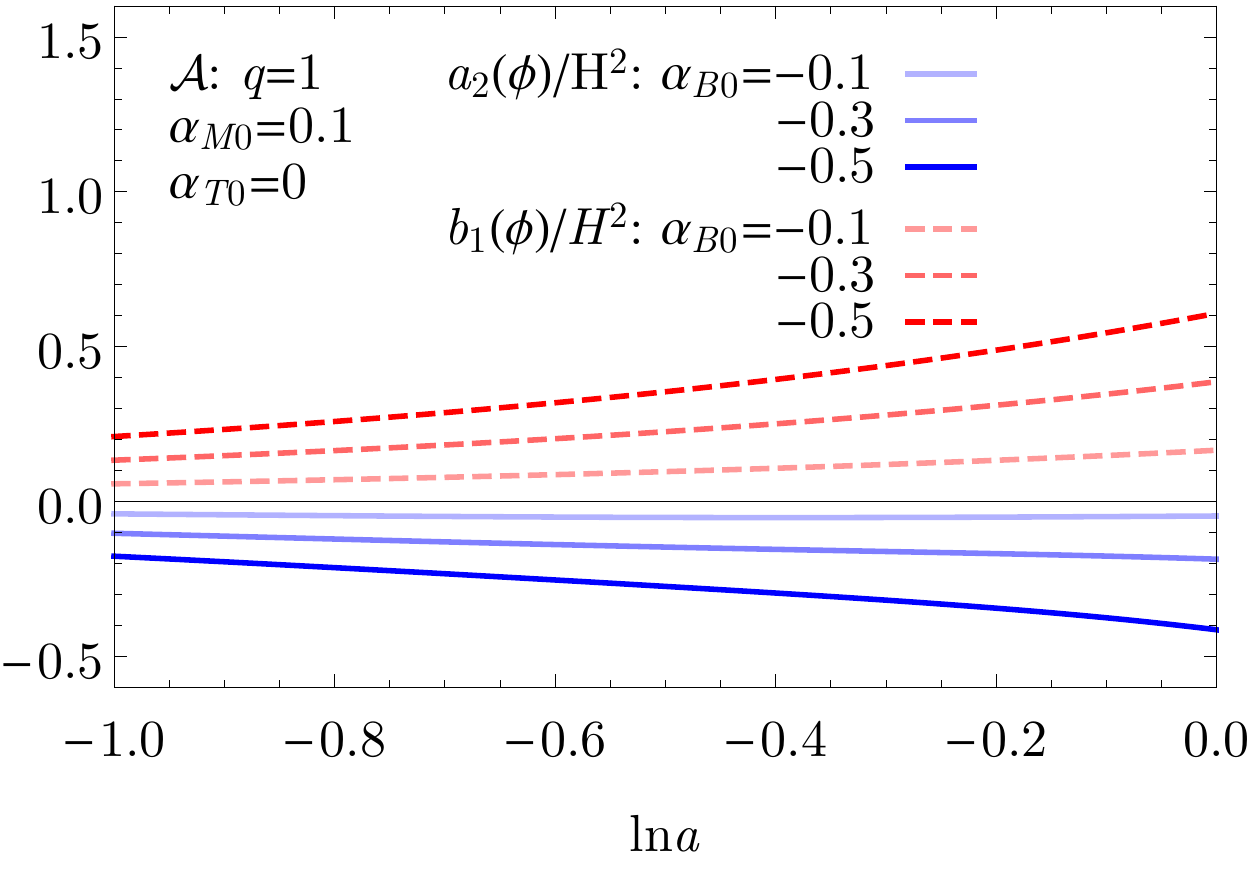}
}
\label{Comparplot1}\label{Comparplot2}\label{Comparplot3}\label{Comparplot4}
\caption{Effects on the reconstructed scalar-tensor theory from
incremental changes in the amplitude of $\alpha_{M}$ for a fixed $\alpha_{B}$ and vice versa.
The general form of the underlying theory is rather insensitive to these changes.
Enhancing $\alpha_{B}$ suppresses the potential and enhances all the other terms whereas enhancing $\alpha_M$  increases every term in the reconstruction other than the k-essence term $a_{2}(\phi)$.
Note that the color scheme here bears no distinction between dark energy and modified gravity in contrast to
all other figures.}
\label{Amplitudechanges}
\end{figure*}
%%% FIGURE APP 3 %%%

%%% TABLE APP %%%
\begin{table*}[t]
\centering
\begin{tabular}{|c|c|} 
\hline
\multicolumn{2}{|c|}{$U(\phi) = \Lambda + \frac{\Gamma}{2} - \frac{M_{2}^{4}}{2M_{*}^{2}} - \frac{9H\bar{M}_{1}^{3}}{8M_{*}^{2}} - \frac{(\bar{M}_{1}^{3})^{\prime}}{8}+\frac{M_{*}^{2}(\bar{M}^{2}_{2})^{\prime\prime}}{4}+\frac{7(\bar{M}_{2}^{2})^{\prime}H}{4}+\bar{M}_{2}^{2}H^{\prime}+\frac{9H^{2}\bar{M}_{2}^{2}}{2M_{*}^{2}}$} \Tstrut \Bstrut \\ 
\hline
\multicolumn{2}{|c|}{$Z(\phi) = \frac{\Gamma}{M_{*}^{4}} - \frac{2M_{2}^{4}}{M_{*}^{6}} - \frac{3H\bar{M}_{1}^{3}}{2M_{*}^{6}} + \frac{(\bar{M}_{1}^{3})^{\prime}}{2M_{*}^{4}}-\frac{(\bar{M}_{2}^{2})^{\prime\prime}}{M_{*}^{2}}-\frac{H(\bar{M}_{2}^{2})^{\prime}}{M_{*}^{4}}-\frac{4H^{\prime}\bar{M}_{2}^{2}}{M_{*}^{4}}$} \Tstrut \Bstrut  \\ \hline 
\multicolumn{2}{|c|}
{$a_{2}(\phi)=\frac{M_{2}^{4}}{2M_{*}^{8}}+\frac{(\bar{M}^{3}_{1})^{\prime}}{8M_{*}^{6}}-\frac{3H\bar{M}_{1}^{3}}{8M_{*}^{8}}-\frac{(\bar{M}_{2}^{2})^{\prime\prime}}{4M_{*}^{4}}+\frac{H(\bar{M}_{2}^{2})^{\prime}}{4M_{*}^{6}}+\frac{H^{\prime}\bar{M}_{2}^{2}}{M_{*}^{6}}-\frac{3H^{2}\bar{M}_{2}^{2}}{2M_{*}^{8}}$} \Bstrut \Tstrut \\ 
\hline                   
\hspace{1.9cm}$ b_{0}(\phi)=0 $ \hspace{1.9cm}  &  $b_{1}(\phi)=\frac{2H\bar{M}_{2}^{2}}{M_{*}^{6}}-\frac{(\bar{M}_{2}^{2})^{\prime}}{M_{*}^{4}}+\frac{\bar{M}_{1}^{3}}{2M_{*}^{6}} $  \Bstrut \Tstrut  \\ 
\hline 
$F(\phi)=\Omega+\frac{\bar{M}_{2}^{2}}{M_{*}^{2}}$ & $c_1(\phi)=\frac{\bar{M}_{2}^{2}}{2M_{*}^{4}}$ \Bstrut \Tstrut \\ 
\hline
\end{tabular}
\caption{\label{solution} The various contributions to the Horndeski functions $G_i(\phi,X)$ in Eqs.~\eqref{eq:G2recon}--\eqref{eq:G4recon}, arising from the reconstruction of the EFT functions of the unitary gauge action in Eqs~\eqref{eq:s01} and \eqref{eq:s2} (see Ref.~\cite{Kennedy:2017sof}).
} 
\end{table*}
%%% TABLE APP %%%

Finally, we examine the sensitivity of the reconstructed theories on the variation of parameter values for a given parametrization of the EFT functions.
We shall only use the functional form $\mathcal{A}$, discussed in Sec.~\ref{sec:IIIA}, which is broadly used in literature.
Recall that we have found that the form of the underlying theory is rather insensitive to the choice between functions $\mathcal{A}$ and $\mathcal{B}$ (Fig.~\ref{Comparisonofparams}).
In all cases we check that the stability condition $\alpha>0$ is satisfied and with the remaining freedom in $\alpha_{K}$ we set $c_{s}^{2}=1$.
We furthermore set $\alpha_{T}=0$.
As a consequence of these choices, the signs of $\alpha_{B}$ and $\alpha_{M}$ are opposite.

In Fig.~\ref{Enhancedplots} we show the effect on the theory when the braiding term $\alpha_{B}$ dominates over the variation in the Planck mass $\alpha_{M}$ and vice versa.
In the first instance, the dominant terms are a potential behaving like a cosmological constant and a large kinetic term for the scalar field mimicking a Brans-Dicke theory with small k-essence and cubic Galileon contributions.
On the contrary, when $\alpha_{B}$ dominates over $\alpha_{M}$ the cubic term $b_{1}(\phi)$ becomes the most relevant term in the theory with the potential decaying away rapidly towards $z=0$.
In both scenarios the $\Lambda $CDM expansion history is maintained by the behavior of the complementary terms in the reconstruction that compensate for the change in the potential.

Next, we examine the effects of varying the power in the parametrization
while retaining consistency in
the stability requirements.
We fix the magnitude of $\alpha_{M0}$ and $\alpha_{B0}$ to be equal but opposite.
The effects of changing the power on the underlying theory are illustrated in Fig.~\ref{VaryingSlope}.
When the power of the parameterization is increased
the effects of modified gravity become more relevant at later times.
The cubic term is generally unaffected by this variation, but the kinetic and k-essence terms are enhanced.
When a large power is chosen, the k-essence
contribution
comes to dominate at late times.  

Finally, in Fig.~\ref{Amplitudechanges} we
illustrate the effects
of changing $\alpha_{M0}$ while keeping $\alpha_{B0}$ fixed and vice versa.
We find that the form of the underlying theory is fairly insensitive to small changes in the amplitude, although certain terms may be enhanced or suppressed relative to others with different choices.
For example, increasing $\alpha_{M}$ has the effect of enhancing the potential relative to that of $\Lambda$CDM.
This is again due to the dependence of $\Lambda \sim M^{2}$.
The kinetic term $Z(\phi)$ is also enhanced although to a lesser degree than the potential whereas the k-essence and cubic Galileon terms $a_{2}(\phi)$ and $b_{1}(\phi)$ are rather insensitive to these $\mathcal{O}(10^{-1})$ changes in $\alpha_{M}$.
The term $a_{2}(\phi)$
remains least affected with smaller variations restricted to the past.
Thus, in general we find that by enhancing $\alpha_{M0}$ for a fixed, small $\alpha_{B0}$, one is enhancing the potential and the standard kinetic term of the scalar-tensor model.
In contrast, for a fixed small value of $\alpha_{M0}$, enhancing the effects of $\alpha_{B0}$ leads to a suppression of the potential and an enhancement of the cubic Galileon term.

%%%% ACKNOWLEDGMENTS %%%%

\acknowledgments

This work is supported by the STFC Consolidated Grant for Astronomy and Astrophysics at the University of Edinburgh. J.K.~thanks STFC for support through an STFC studentship. L.L.~also acknowledges support by a Swiss National Science Foundation Professorship grant (No.~170547) and Advanced Postdoc.Mobility Fellowship
(No.~161058). A.N.T.~thanks the Royal Society for support from a Wolfson Research Merit Award. Please contact the authors for access to research materials.

%%%%%% BIBLIOGRAHY %%%%%%
\clearpage
\bibliographystyle{JHEP}
\bibliography{library}
%%%%%%%%%%%%%%%%%%%%%%%%

\end{document}